\documentclass[reprint,showpacs,prd,preprintnumbers,amssymb,nofootinbib,superscriptaddress,aps,onecolumn]{revtex4-2}
\usepackage{graphicx,floatflt,amssymb,rotate,epsfig}
\usepackage[utf8]{inputenc}
\usepackage[inline]{enumitem}
\usepackage{mathrsfs}
\usepackage{amssymb}
\usepackage{amsmath}
\usepackage{color}
\usepackage{graphicx}
\usepackage{subfig}
\usepackage{multirow}
\usepackage{mathtools}
\usepackage{float}
\usepackage{pstricks}
\usepackage[toc,page]{appendix}
\usepackage{pifont}
\usepackage{braket}
\usepackage{bbold}
\usepackage{hyperref}

\usepackage{relsize}
\def\Babar{{\mbox{\slshape B\kern-0.1em{\smaller A}\kern-0.1em B\kern-0.1em{\smaller A\kern-0.2em R}}}}

\newcommand{\ba}{\begin{array}}
	\newcommand{\ea}{\end{array}}
\def\beq{\begin{equation}}
\def\eeq{\end{equation}}
\def\bea{\begin{eqnarray}}
\def\eea{\end{eqnarray}}
\def\nn{\nonumber}

\def\roughly#1{\mathrel{\raise.3ex\hbox
		{$#1$\kern-.75em\lower1ex\hbox{$\sim$}}}}

\def\sla#1{\raise.15ex\hbox{$/$}\kern-.57em #1}

\def\order{\lower 1.8ex \hbox{\LARGE\~{}}}

\def\btopilnu{B \to \pi \ell \nu_{\ell}}
\def\btorholnu{B \to \rho \ell \nu_{\ell}}
\def\btopitaunu{B \to \pi \tau \nu_{\tau}}
\def\btorhotaunu{B \to \rho \tau \nu_{\tau}}

\def\btotaunu{B \to \tau \nu_{\tau}}

\newcommand*{\rom}[1]{\expandafter\@slowromancap\romannumeral #1@}

\newcommand{\Bbar}{\,\bar{\!B}}

\def\Bbar{\overline{B}}

\def\ubar{\overline{u}}

\def\taubar{\overline{\tau}}

\def\nubar{{\overline{\nu}}}

\def\Heff{\mathcal{H}_{\rm eff}}

\def\O{\mathcal{O}}

\def\Re{\mathcal{R}e}

\def\Dst{{D^*}}

\usepackage[normalem]{ulem}

\definecolor{darkgreen}{cmyk}{1,0,1,0.4}

\definecolor{pink}{cmyk}{0.4,1,0.3,0}

\def\com2#1{\textcolor{red}{\it{#1}}}

\begin{document}
	
	\title{A closer look at observables from exclusive semileptonic $B\to(\pi,\rho)\ell\nu_{\ell}$ decays }
	
	\author{Aritra Biswas}
	\email{iluvnpur@gmail.com}
	\affiliation{Indian Institute of Technology, North Guwahati, Guwahati 781039, Assam, India }
	
	\author{Soumitra Nandi}
	\email{soumitra.nandi@iitg.ac.in}
	\affiliation{Indian Institute of Technology, North Guwahati, Guwahati 781039, Assam, India }

\begin{abstract}  
This article analyses the available inputs in $\btopilnu$ and $\btorholnu$ decays which include the measured values of differential rate in different $q^2$-bins (lepton invariant mass spectrum), lattice, and the newly available inputs on the relevant form-factors from the light-cone sum rules (LCSR) approach. We define different fit scenarios, and in each of these scenarios, we predict a few observables in the standard model (SM). For example, $R(M) =\frac{\mathcal{B}(B \to M\ell_i\nu_{\ell_i})}{\mathcal{B}(B\to M\ell_j\nu_{\ell_j})} $, $R^{\ell_i}_{\ell_j}(M) =\frac{\mathcal{B}(B\to \ell_i\nu_{\ell_i})}{\mathcal{B}(B \to M\ell_j\nu_{\ell_j})}$ with  M = $\pi$ or $\rho$ and $\ell_{i,j} = e, \mu$ or $\tau$. We also discuss the new physics (NP) sensitivities of all these observables and obtain bounds on a few NP Wilson coefficients in $b\to u \tau\nu_{\tau}$ decays using the available data. We have noted that the data at present allows sizeable NP contributions in this mode. Also, we have predicted a few angular observables relevant to these decay modes.

\end{abstract}   
	
	\maketitle
\section{Introduction}

The exclusive decays like $\btopilnu$ and $\btorholnu$ ($\ell=\mu$ or $e$) are used to extract the Cabibbo-Kobayashi-Maskawa (CKM) matrix element $|V_{ub}|$; for details see \cite{Amhis:2019ckw} and the references therein. For a recent update, see refs. \cite{Leljak:2021vte,Biswas:2021qyq,Bernlochner:2021rel}. Within the SM, the decay modes mentioned above are mediated by a tree-level charged current interaction. The general expectation is that these decay modes with the $\mu$ and $e$ in the final state are insensitive to any NP effects. However, the decay modes $\btopitaunu$ and $\btorhotaunu$ could be sensitive to new fundamental interactions beyond the SM (BSM) due to a relatively large mass of $\tau$. As a reference one could look at some eralier studies on these channels \cite{Kang:2018jzg,Colangelo:2019axi,Zhang:2020dla,Becirevic:2020rzi,Fleischer:2021yjo}. On a similar note the observables associated with the purely leptonic decay $\btotaunu$ could also be sensitive to new interactions.     

We encounter a similar situation in $b\to c\ell\nu_{\ell}$ decays. The decay modes with $\mu$ or $e$ in the final states are used to extract $|V_{cb}|$, while the ratios $R(D^{(*)}) = \frac{\Gamma(B\to D^{(*)}\tau\nu_{\tau})}{\Gamma(B\to D^{(*)}\ell\nu_{\ell})}$ (with $\ell=\mu$ or $e$) are sensitive to lepton-universality-violating (LUV) NP effects in these decay modes, for an update see \cite{hflavWeb,Jaiswal:2017rve,Jaiswal:2020wer}. In accordance with these observables, one can also define the following:
\begin{equation}
R(M) =\frac{\mathcal{B}(B \to M\tau\nu_{\tau})}{\mathcal{B}(B\to M\ell\nu_{\ell})},\ \ R^{\tau}_{\ell}(M) =\frac{\mathcal{B}(B\to \tau\nu_{\tau})}{\mathcal{B}(B \to M\ell\nu_{\ell})},\ \ R^{\mu}_{\tau}(M) =\frac{\mathcal{B}(B\to \mu\nu_{\mu})}{\mathcal{B}(B \to M\tau\nu_{\tau})},\ \ R^{\tau}_{\tau}(M) =\frac{\mathcal{B}(B\to \tau\nu_{\tau})}{\mathcal{B}(B \to M \tau\nu_{\tau})}\ ,
\label{eq:obs}
\end{equation}
here $M$ is either $\pi$ or $\rho$ while, as before, $\ell$ is muon or electron. In an earlier analysis \cite{Banelli:2018fnx}, a few of the above observables has been mentioned, here we will carry out a thorough analysis based on the newly available information. The NP sensitivities of these observables may be complementary to each other and might help distinguish different beyond the SM (BSM) interactions. One of the major sources of error in the predictions of $\mathcal{B}(B\to M\tau\nu_{\tau})$ or $\mathcal{B}(B\to \tau\nu_{\tau})$ decays is the CKM element $|V_{ub}|$ which cancels in the ratios as mentioned above. The only source of uncertainties, therefore, are the form-factors. There are two form-factors associated with $\btopilnu$ decays, namely $f_+(q^2)$ and  $f_0(q^2)$\footnote{The sensitivity to $f_0(q^2)$ is suppressed for lighter leptons.}, for which precise predictions from lattice at zero and non-zero recoils are available \cite{Flynn:2015mha,Lattice:2015tia}, while the updates from LCSR is available in \cite{Gubernari:2018wyi,Leljak:2021vte}. The analysis in \cite{Leljak:2021vte} uses the two-particle twist-two pion light-cone distribution amplitude (LCDA), and the results are more precise than those obtained in \cite{Gubernari:2018wyi} which is an LO calculation with the ill-known B-meson LCDA. We have analysed the data including the results from both these publications. For $\btopitaunu$ decays there will be an additional form-factor $f_T(q^2)$ assuming NP affects only tauonic final states. The inputs on $f_T$ are available from lattice \cite{Bailey:2015nbd}, as well as from LCSR \cite{Gubernari:2018wyi,Leljak:2021vte}. On top of these, we have inputs on the differential branching fractions in different $q^2$-bins \cite{Ha:2010rf, Lees:2012vv, Sibidanov:2013rkk} which play an essential role in constraining the form-factors. On the other hand, the $\btorholnu$ decays involve four form-factors. Three of them, $A_1(q^2)$, $A_2(q^2)$, $A_0(q^2)$ are associated with the axial-vector current and the one obtained from the vector current is $V(q^2)$. In these decays, we do not have any information from lattice. We have a few inputs from LCSR on each of these form-factors \cite{Straub:2015ica,Gubernari:2018wyi}. The LCSR results in ref.~\cite{Straub:2015ica} have been derived up to twist-3 $\mathcal{O}(\alpha_s)$ using the $\rho$ meson LCDA, and the extracted values are relatively more precise than the ones obtained in \cite{Gubernari:2018wyi}. In ref. \cite{Gubernari:2018wyi} the computation of the $B\to\rho$ form factors are done within the narrow-width approximation of the $\rho$ meson. However, the $\rho$-meson is an unstable particle and decays strongly to pairs of pseudoscalar mesons. Therefore, both theory and experimental analyses of $B\to\rho\ell\nu_{\ell}$ will be sensitive to the treatment of the background, finite width and S, P-wave effects. From the experimental point of view, $\rho$-meson is detected as a Breit-Wigner peak in the invariant mass distribution of produced pions ($\rho\to \pi\pi$). In ref. \cite{Straub:2015ica}, the computation goes beyond the narrow-width approximation. The $\rho$ meson DA is characterized by the longitudinal and transverse component of the decay constant $f_{\rho}^{\parallel}$ and $f_{\rho}^{\perp}$, respectively. The inputs on $f_{\rho}^{\parallel}$ are obtained from the measurements of the decay widths: $\Gamma(e^+e^- \to \rho^0(\to \pi\pi))$ and $\Gamma(\tau^{+} \to \rho^{+}(\pi\pi)\nu)$ respectively \cite{pdgrev}. In those analyses, the amplitudes of the Breit-Wigner ansatz of the resonant $\rho$, $\rho(1450)$, and $\rho(1700)$ states are fitted from the $\pi\pi$ spectrum in a certain mass window around $ m_{\pi\pi}^2 \approx m_{\rho}^2 $. While the transverse component $f^{\perp}_{\rho}$ is obtained from the lattice estimates of the ratio $f_{\rho}^{\parallel}/f^{\perp}_{\rho}$. As argued in \cite{Straub:2015ica}, as long as the treatment of the $\rho (\to \pi\pi)$ meson is the same as is used in the experimental extractions of $f_{\rho}^{\parallel}$, there is no systematic effect. The LCSR should not suffer from sizeable additional uncertainties. Data is also available on the differential rates in different $q^2$-bins \cite{Lees:2012vv}. 

Our primary goal is to predict the observables given in eq. \ref{eq:obs} first in the SM and then in the specific NP scenarios wherever applicable. In addition, we will predict a few more observables in $B\to \rho\ell\nu_{\ell}$ decays, which we will discuss later. To do so, we need to constrain the form-factors (discussed above) over all of the allowed $q^2$ region using the available pieces of information from the lattice, LCSR and experimental data. For a comparative study, we have defined different fit scenarios. In each of these fit scenarios, we have estimated the best fit values and the respective errors for different observables (provided above) in the SM. Due to helicity suppression, the charged current decay $\btotaunu$ is potentially sensitive to BSM interactions. At the moment, one of the major sources of uncertainties in the SM predictions for the branching fraction $\mathcal{B}(\btotaunu)$ is the CKM element $|V_{ub}|$. For one of the fit scenarios, we have created synthetic data points corresponding to the ratio $R^{\tau}_{\ell}(M)$. In $R^{\tau}_{\ell}(M)$ the CKM element $|V_{ub}|$ gets cancelled, and these observables are useful to simultaneously constrain the form-factor parameters and the new couplings. Note that experimental data is available on the branching fraction $\mathcal{B}(\btotaunu)$ but not on $\mathcal{B}(B\to \mu\nu_{\mu})$\footnote{PDG metnions a range of $2.9\times 10^{-7}$ to $1.07\times 10^{-6}$ at $90\%$ CL for $\mathcal{B}(B\to\mu\nu)$.}. Using these fit results, we have given several predictions in the SM and the NP scenarios. Following that, we have studied the NP sensitivities of different observables, as mentioned above.

\section{Theory}\label{sec:theory}
Assuming neutrinos to be left-handed,the most general effective Hamiltonian that contains all possible four-fermion operators of the lowest dimension for the $b\to u \tau\nubar$ transition\footnote{We ommit the lepton index from the New Physics (NP) Wilson coefficients (WC's) since we assume that the NP effects operators with tauonic final states only.} is written as,
\begin{equation}
\Heff = {4G_F \over \sqrt2} V_{ub}\left[ (1 + C_{V_1})\O_{V_1} + C_{V_2}\O_{V_2} + C_{S_1}\O_{S_1} + C_{S_2}\O_{S_2} + C_T\O_T \right] \,,
\label{eq:Heff}
\end{equation}
with the operator basis defined as
\begin{equation}
\begin{split}
\O_{V_1} =& (\ubar_L \gamma^\mu b_L)(\taubar_L \gamma_\mu \nu_{L}) \,, \\
\O_{V_2} =& (\ubar_R \gamma^\mu b_R)(\taubar_L \gamma_\mu \nu_{L}) \,, \\
\O_{S_1} =& (\ubar_L b_R)(\taubar_R \nu_{L}) \,, \\
\O_{S_2} =& (\ubar_R b_L)(\taubar_R \nu_{L}) \,, \\
\O_T =& (\ubar_R \sigma^{\mu\nu} b_L)(\taubar_R \sigma_{\mu\nu} \nu_{L}) \,.
\end{split}
\label{eq:operators}
\end{equation}
\footnote{Since neutrinos are known to undergo mixing, we omit the lepton index over neutrinos.} In the SM, the Wilson coefficients are set to zero, $C_X=0$ ($X=V_{1,2},\,S_{1,2},\,T$).
 
For the above Hamiltonian, the semileptonic decay width distribution for $B\to\pi\tau\nu$ transitions is:
\begin{equation}\label{eq:brsemileppi}
\begin{split}
{d\Gamma(\Bbar \to \pi\tau\nubar) \over dq^2} =& {G_F^2 |V_{ub}|^2 \over 192\pi^3 m_B^3} q^2 \sqrt{\lambda_\pi(q^2)} \left( 1 - {m_\tau^2 \over q^2} \right)^2 \times\biggl\{ |1 + C_{V_1} + C_{V_2}|^2 \left[ \left( 1 + {m_\tau^2 \over2q^2} \right) H_{V,0}^{s\,2} + {3 \over 2}{m_\tau^2 \over q^2} \, H_{V,t}^{s\,2} \right] \\
&+ {3 \over 2} |C_S|^2 \, H_S^{s\,2} + 8|C_T|^2 \left( 1+ {2m_\tau^2 \over q^2} \right) \, H_T^{s\,2}+ 3\Re[ ( 1 + C_{V_1} + C_{V_2} ) (C_S^{*} ) ] {m_\tau \over \sqrt{q^2}} \, H_S^s H_{V,t}^s \\
&- 12\Re[ ( 1 + C_{V_1} + C_{V_2} ) C_T^{*} ] {m_\tau \over \sqrt{q^2}} \, H_T^s H_{V,0}^s \biggl.\biggr\}
\end{split}
\end{equation}
where $\lambda_\pi = ((m_B-m_\pi)^2-q^2)((m_B+m_\pi)^2-q^2)$, $C_S = C_{S_1} +C_{S_2}$, and
\begin{subequations}
	\begin{align}
	H_{V,0}^s(q^2) \equiv& \, H_{V_1,0}^s(q^2) = H_{V_2,0}^s(q^2) = \sqrt{\lambda_\pi(q^2) \over q^2} f_+(q^2) \,, \\
	& \nonumber \\
	H_{V,t}^s(q^2) \equiv& \, H_{V_1,t}^s(q^2) = H_{V_2,t}^s(q^2) = {m_B^2-m_\pi^2 \over \sqrt{q^2}} f_0(q^2) \,, \\
	& \nonumber \\
	H_S^s(q^2) \equiv& \, H_{S_1}^s(q^2) = H_{S_2}^s(q^2) \simeq {m_B^2-m_\pi^2 \over m_b-m_u} f_0(q^2) \,, \\
	& \nonumber \\
	H_T^s(q^2) \equiv& \, H_{T,+-}^s(q^2) = H_{T,0t}^s(q^2) = -{\sqrt{\lambda_\pi(q^2)} \over m_B+m_\pi} f_T(q^2).
	\end{align}
\end{subequations}
The semileptonic decay width distribution for $B\to\rho\tau\nu$ transitions is written as:
\begin{equation}\label{eq:brsemileprho}
\begin{split}
{d\Gamma(\Bbar \to \rho \tau\nubar) \over dq^2} = & {G_F^2 |V_{cb}|^2 \over 192\pi^3 m_B^3} q^2 \sqrt{\lambda_\rho(q^2)} \left( 1 - {m_\tau^2 \over q^2} \right)^2 \times\biggl\{ \biggr.   ( |1 + C_{V_1}|^2 + |C_{V_2}|^2 ) \left[ \left( 1 + {m_\tau^2 \over2q^2} \right) \left( H_{V,+}^2 + H_{V,-}^2 + H_{V,0}^2 \right)\right. \\
&\left.+ {3 \over 2}{m_\tau^2 \over q^2} \, H_{V,t}^2 \right] - 2\Re[(1 + C_{V_1}) C_{V_2}^{*}] \left[ \left( 1 + {m_\tau^2 \over 2q^2} \right) \left( H_{V,0}^2 + 2 H_{V,+} H_{V,-} \right) + {3 \over 2}{m_\tau^2 \over q^2} \, H_{V,t}^2 \right] \\
& + {3 \over 2} |C_P|^2 \,H_S^2  + 8|C_T|^2 \left( 1+ {2m_\tau^2 \over q^2} \right) \left( H_{T,+}^2 + H_{T,-}^2 + H_{T,0}^2  \right)  + 3\Re[ ( 1 + C_{V_1} - C_{V_2} )\\
& C_P^{*} ] {m_\tau \over \sqrt{q^2}} \, H_S H_{V,t} - 12\Re[ (1 + C_{V_1}) C_T^{*} ] {m_\tau \over \sqrt{q^2}} \left( H_{T,0} H_{V,0} + H_{T,+} H_{V,+} - H_{T,-} H_{V,-} \right)\\
& + 12\Re[ C_{V_2} C_T^{*} ] {m_\tau \over \sqrt{q^2}} \left( H_{T,0} H_{V,0} + H_{T,+} H_{V,-} - H_{T,-} H_{V,+} \right) \biggl.\biggr\} 
\end{split}
\end{equation}
where $\lambda_\rho = ((m_B-m_\rho)^2-q^2)((m_B+m_\rho)^2-q^2)$, $C_P = C_{S_1} - C_{S_2}$, and
\begin{subequations}
	\begin{align}
	H_{V,\pm}(q^2) \equiv& \, H_{V_1,\pm}^\pm(q^2) = -H_{V_2,\mp}^\mp(q^2) = (m_B+m_\Dst) A_1(q^2) \mp { \sqrt{\lambda_\Dst(q^2)} \over m_B+m_\Dst } V(q^2) \,, \\
	& \nonumber \\
	H_{V,0}(q^2) \equiv& \, H_{V_1,0}^0(q^2) = -H_{V_2,0}^0(q^2) = { m_B+m_\Dst \over 2m_\Dst\sqrt{q^2} } \left[ -(m_B^2-m_\Dst^2-q^2) A_1(q^2) \right.\left. + { \lambda_\Dst(q^2) \over (m_B+m_\Dst)^2 } A_2(q^2) \right] \,, \\
	& \nonumber \\
	H_{V,t}(q^2) \equiv& \, H_{V_1,t}^0(q^2) = -H_{V_2,t}^0(q^2) = -\sqrt{ \lambda_\Dst(q^2) \over q^2 } A_0(q^2) \,, \\
	& \nonumber \\
	H_S(q^2) \equiv& \, H_{S_1}^0(q^2) = -H_{S_2}^0(q^2) \simeq -{ \sqrt{\lambda_\Dst(q^2)} \over m_b+m_c } A_0(q^2) \,, \\
	& \nonumber \\
	H_{T,\pm}(q^2) \equiv& \, \pm H_{T,\pm t}^\pm(q^2) = { 1 \over \sqrt{q^2} } \left[ \pm (m_B^2-m_\Dst^2) T_2(q^2) + \sqrt{\lambda_\Dst(q^2)} T_1(q^2) \right] \,, \\
	& \nonumber \\
	H_{T,0}(q^2) \equiv& \, H_{T,+-}^0(q^2) = H_{T,0t}^0(q^2) = { 1 \over 2m_\Dst } \left[ -(m_B^2+3m_\Dst^2-q^2) T_2(q^2) \right.\left. + { \lambda_\Dst(q^2) \over m_B^2-m_\Dst^2 } T_3(q^2) \right] \,.
	\end{align}
\end{subequations}
  
The branching fraction for $B\to\tau\nu$ corresponding to the same Hamiltonian is:
\begin{equation}\label{eq:brlep}
\begin{split}
\mathcal{B}(B\to\tau\nu) = & \frac{\tau_B}{8\pi}m_Bm_\tau f_B^2G_f^2V_{ub}^2(1-\frac{m_\tau^2}{m_B^2})\left|1+C_{V_1}-C_{V_2}+\frac{m_B^2}{m_\tau(m_b+m_u)}C_P\right|^2.
\end{split}
\end{equation}
The respective branching fractions for $B\to\mu (e)\nu$ can be obtained by replacing $m_{\tau} \rightarrow {m_{\mu}}(m_e)$ in eq. \ref{eq:brlep}.  In the SM, the predicted value is given by 
\begin{equation}
\mathcal{B}(B\to\tau\nu)_{SM} = (0.975 \pm 0.068)\times 10^{-4}.
\end{equation}
In order to obtain the SM prediction, we have used $f_B=0.19\pm 0.0013$ which is the $N_f=2+1+1$ FLAG 2019 average~\cite{FlavourLatticeAveragingGroup:2019iem} and $|V_{ub}| = (3.91 \pm 0.13)\times 10^{-3}$. This value has been obtained as a result of updating the value of $V_{ub} = (3.93 \pm 0.14)\times 10^{-3}$ reported in our previous publication~\cite{Biswas:2021qyq} after the inclusion of the $B\to\pi$ LCSR data provided in~\cite{Leljak:2021vte}.

Note that the decay rates for $B\to \pi\tau\nu$ and $B \to \rho\tau\nu$ are sensitive to both the $(V + A)$ and $(V - A)$ type of quark currents and tensor type interaction $\mathcal{O}_T$. Interestingly, subject to the scalar (S) and pseudoscalar (P) type of interactions, the decay rate $\Gamma(B\to \pi\tau \nu)$ is sensitive only to S-type interaction while the rate $\Gamma(B\to \rho\tau \nu)$ is sensitive to $P$-type interaction only. On the other hand, the braching fraction $\mathcal{B}(B\to \tau\nu)$ is sensitive to $(V\pm A)$ and P-type quark currents. However, it is insensitive to scalar and tensor interactions. With all this information at hand, we define the following sets of observables: 
\begin{eqnarray}
R(\pi) &=&\frac{\Gamma(B\to\pi\tau\nu)}{\Gamma(B\to\pi\mu\nu)},\ \ R_{\mu}^\tau(\pi) =\frac{\Gamma(B\to\tau\nu)}{\Gamma(B\to\pi\mu\nu)},\ \  R_{\tau}^\tau(\pi) = \frac{\Gamma(B\to\tau\nu)}{\Gamma(B\to\pi\tau\nu)}, \ \ R_{\tau}^\mu(\pi) = \frac{\Gamma(B\to\mu\nu)}{\Gamma(B\to\pi\tau\nu)}\label{eq:Rpiobs}
\end{eqnarray}
\begin{eqnarray}
R(\rho) &=&\frac{\Gamma(B\to\rho\tau\nu)}{\Gamma(B\to\rho\mu\nu)},\ \ R_{\mu}^\tau(\rho) =\frac{\Gamma(B\to\tau\nu)}{\Gamma(B\to\rho\mu\nu)},\ \  R_{\tau}^\tau(\rho) = \frac{\Gamma(B\to\tau\nu)}{\Gamma(B\to\rho\tau\nu)}, \ \ R_{\tau}^\mu(\rho) = \frac{\Gamma(B\to\mu\nu)}{\Gamma(B\to\rho\tau\nu)}\label{eq:Rrhoobs}
\end{eqnarray}
Since we are not considering NP effects in $b\to u\ell \nu$ (with $\ell =\mu$ or $e$) decays,  
the NP sensitivities of $R(\pi)$ (or $R_{\tau}^\mu(\pi)$) and $R(\rho)$ (or $R_{\tau}^\mu(\rho)$)  
will be similar to those manifest in $\Gamma(B\to\pi\tau\nu)$ and $\Gamma(B\to\rho\tau\nu)$, respectively.
Moreover, $R_{\mu}^\tau(\pi)$ (or $R_{\mu}^\tau(\rho)$) will display NP effects similar to that visible in $B\to\tau\nu$ decays. For one operator scenarios, the observables defined above could provide the following complementary information:
\begin{itemize}
	\item 
		The contributions from a new SM type interaction will cancel in the ratios $R_{\tau}^\tau(\pi)$ and $R_{\tau}^\tau(\rho)$. Hence, new $(V-A)$-type of interactions can not be probed via these ratios. However, the other ratios as defined in eqs. \ref{eq:Rpiobs} and \ref{eq:Rrhoobs} are sensitive to such interactions. If, in the future, data shows deviations in $R(\pi)$, $R(\rho)$, $R_{\mu}^\tau(\pi)$, and $R_{\mu}^\tau(\rho)$ but not in $R_{\tau}^\tau(\pi)$ and $R_{\tau}^\tau(\rho)$, then that could be an indication of a new SM type interaction.
	
		\item The ratios $R_{\tau}^\tau(\pi)$ and $R_{\tau}^\tau(\rho)$ are sensitive to $\mathcal{O}_{V_2}$, however, the dependences could be very different which we will discuss in the result section. Note that the other observables are also sensitive to $\mathcal{O}_{V_2}$. In the following sections, we will discuss whether we can distinguish these different contributions from each other or not.
	
	\item The observable $R(\rho)$ is sensitive to pseudoscalar current while it is not sensitive to scalar current. In contrast, $R(\pi)$ is sensitive to scalar current, but it is insensitive to pseudoscalar current. Also, $R_{\mu}^\tau(\pi)$, $R_{\mu}^\tau(\rho)$ and $R_{\tau}^\tau(\rho)$ are sensitive to only pseudoscalar current, while $R_{\tau}^\tau(\pi)$ is sensitive to both the scalar and pseudoscalar currents. 
	
	\item Amongst all these observables, $R(\pi)$, $R(\rho)$, $R_{\tau}^\tau(\pi)$, and $R_{\tau}^\tau(\rho)$ are sensitive to tensor currents while $R_{\mu}^\tau(\pi)$ and $R_{\mu}^\tau(\rho)$ are not.         
\end{itemize}
A comparitive study of all these observables in different NP scenarios could be useful to distinguish one scenario from the others, and precise measurements of these observables will be able to probe particular types of interactions. The details will be discussed in the following sections.  

At the moment, there is no measurement on $\mathcal{B}(B\to \pi\tau\nu)$. Only an upper limit is available which is given by $\mathcal{B}(B\to \pi\tau\nu) \le 2.5\times 10^{-4}$~\cite{pdg} at $90\%$ confidence level. The measured value of $\mathcal{B}(B\to \tau\nu) = (1.06 \pm 0.19)\time 10^{-4}$~\cite{pdg}. With these inputs we obtain the following experimental bounds:
\begin{equation}
R(\pi) < 1.6,\ \ \  R^{\tau}_{\mu}(\pi) = 0.71 \pm 0.13,\ \ \  R^{\tau}_{\tau}(\pi) > 0.5. 
\label{eq:expbounds}
\end{equation}  

\section{Analysis and Results}
\begin{table}[t]
	\begin{tabular}{|c|c|c|c|}\hline
		\textbf{Parameters}     & \textbf{Fit-1}  &  \textbf{Fit-2}  &   \textbf{Fit-3}   \\
		&   &  & (No NP in $B\to \tau\nu$) \\\hline
		$\chi ^2\text{/dof}$  &   $\text{18.0939/34}$  &  $\text{75.0986/90}$  & $\text{26.8725/52}$  \\
		$\text{p-value}$  & $0.988372$  &   $0.648645$  &$0.998484$    \\\hline
		$|V_{ub}|$ & - & \text{0.003828(94)} & - \\\hline
		$a_0^+$  &  $\text{0.236(14)}$  &  $\text{0.246(11)}$  &  $\text{0.2416(71)}$  \\
		$a_1^+$  &  $\text{-0.542(98)}$  &  $\text{-0.553(91)}$  &  $\text{-0.605(77)}$  \\
		$a_2^+$  &  $\text{0.48(34)}$  &  $\text{0.36(32)}$  &  $\text{0.18(27)}$  \\
		$a_3^+$  &  $\text{0.82(25)}$  &  $\text{0.73(24)}$  &  $\text{0.60(21)}$  \\
		$a_1^0$  &  $\text{0.477(95)}$  &  $\text{0.529(87)}$  &  $\text{0.497(74)}$  \\
		$a_2^0$  &  $\text{1.49(30)}$  &  $\text{1.58(28)}$  &  $\text{1.50(28)}$  \\
		$a_3^0$  &  $\text{1.72(30)}$  &  $\text{1.78(30)}$  &  $\text{1.71(30)}$  \\
		$a_0^T$  &  $\text{0.231(14)}$  &  $\text{0.239(13)}$  &  $\text{0.236(11)}$  \\
		$a_1^T$  &  $\text{-0.61(11)}$  &  $\text{-0.62(11)}$  &  $\text{-0.64(10)}$  \\
		$a_2^T$  &  $\text{0.15(44)}$  &  $\text{0.021(430)}$  &  $\text{-0.041(412)}$  \\
		$a_3^T$  &  $\text{0.50(40)}$  &  $\text{0.39(39)}$  &  $\text{0.34(38)}$  \\\hline
		\textbf{Observables} & \textbf{Fit-1}  &  \textbf{Fit-2}  &   \textbf{Fit-3}   \\\hline
		$\text{R($\pi $)}$  &  $\text{0.691(15)}$  &  $\text{0.684(11)}$  &  $\text{0.681(13)}$  \\
		$R_{\tau }^{\tau }\text{($\pi $)}$  &  $\text{0.988(65)}$  &  $\text{0.951(41)}$  &  $\text{0.942(53)}$  \\
		$R_{\tau }^{\mu }\text{($\pi $)}$  &  $\text{0.00444(29)}$  &  $\text{0.00427(19)}$  &  $\text{0.00423(24)}$  \\
		$R_{\mu }^{\tau }\text{($\pi $)}$  &  $\text{0.683(55)}$  &  $\text{0.650(31)}$  &  $\text{0.641(43)}$  \\\hline
	\end{tabular}
	\caption{Fit results for the coefficients of the form factors $a_n^i$ for the form factors $f^i(q^2)$ that contribute to semileptonic $B\to\pi$ transitions using a BSZ parametrization as mentioned in the text. The kinematic constraint $f^+(0)=f^0(0)$ is manifest in such a parametrization simply implying that $a_0^+=a_0^0$ and hence $a_0^0$ has not been explicitly shown in the table. Note that the $a_0^T$'s do not take part in semileptonic charged current $B\to\pi$ transitions within the SM and hence are simply constrained by the LCSR and Lattice datapoints.}
	\label{tab:Btopifitresults}
\end{table}

\begin{figure}[ht]
	\centering
	\subfloat[]{\includegraphics[width=0.30\textwidth]{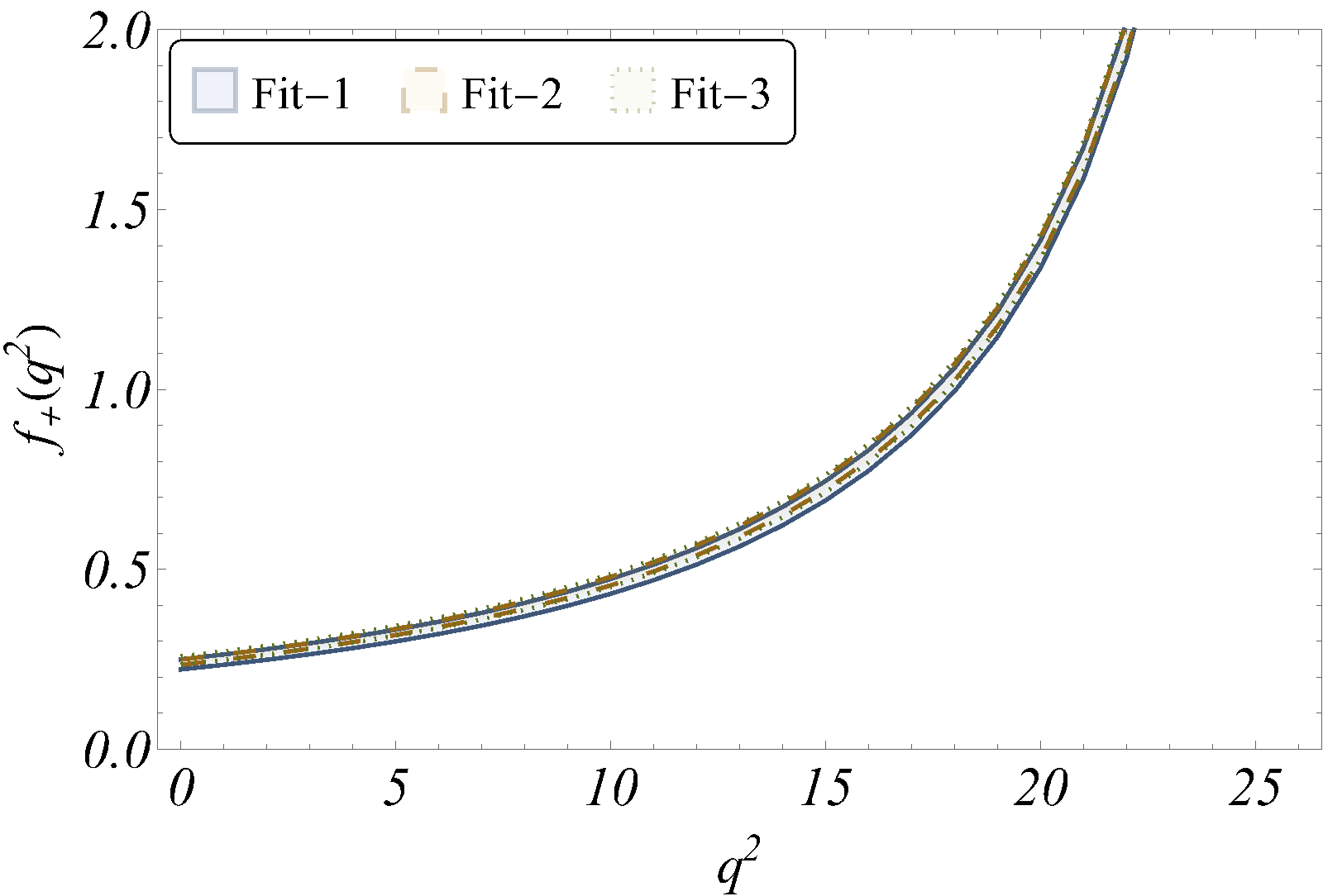}}~
	\subfloat[]{\includegraphics[width=0.30\textwidth]{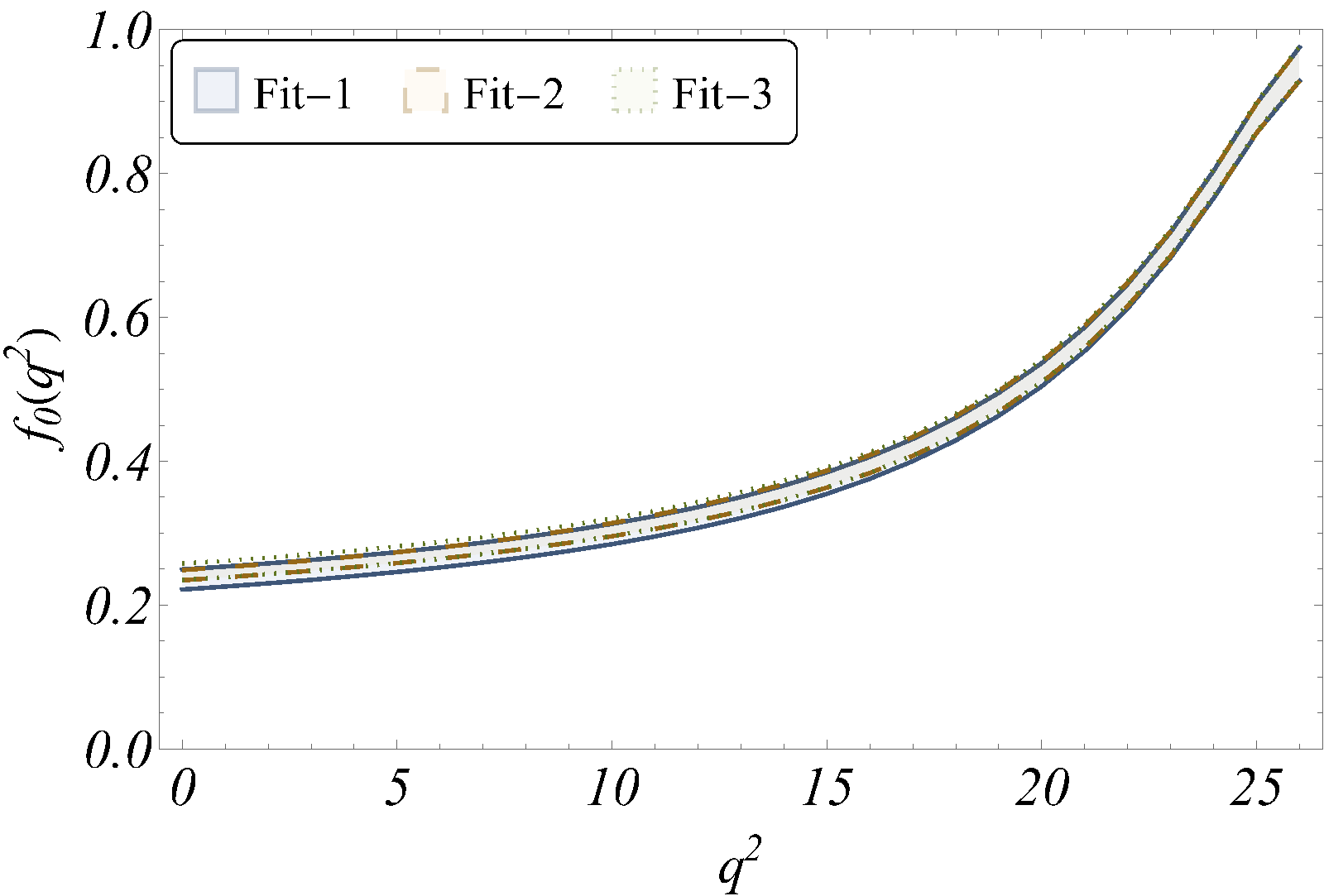}}~
	\subfloat[]{\includegraphics[width=0.30\textwidth]{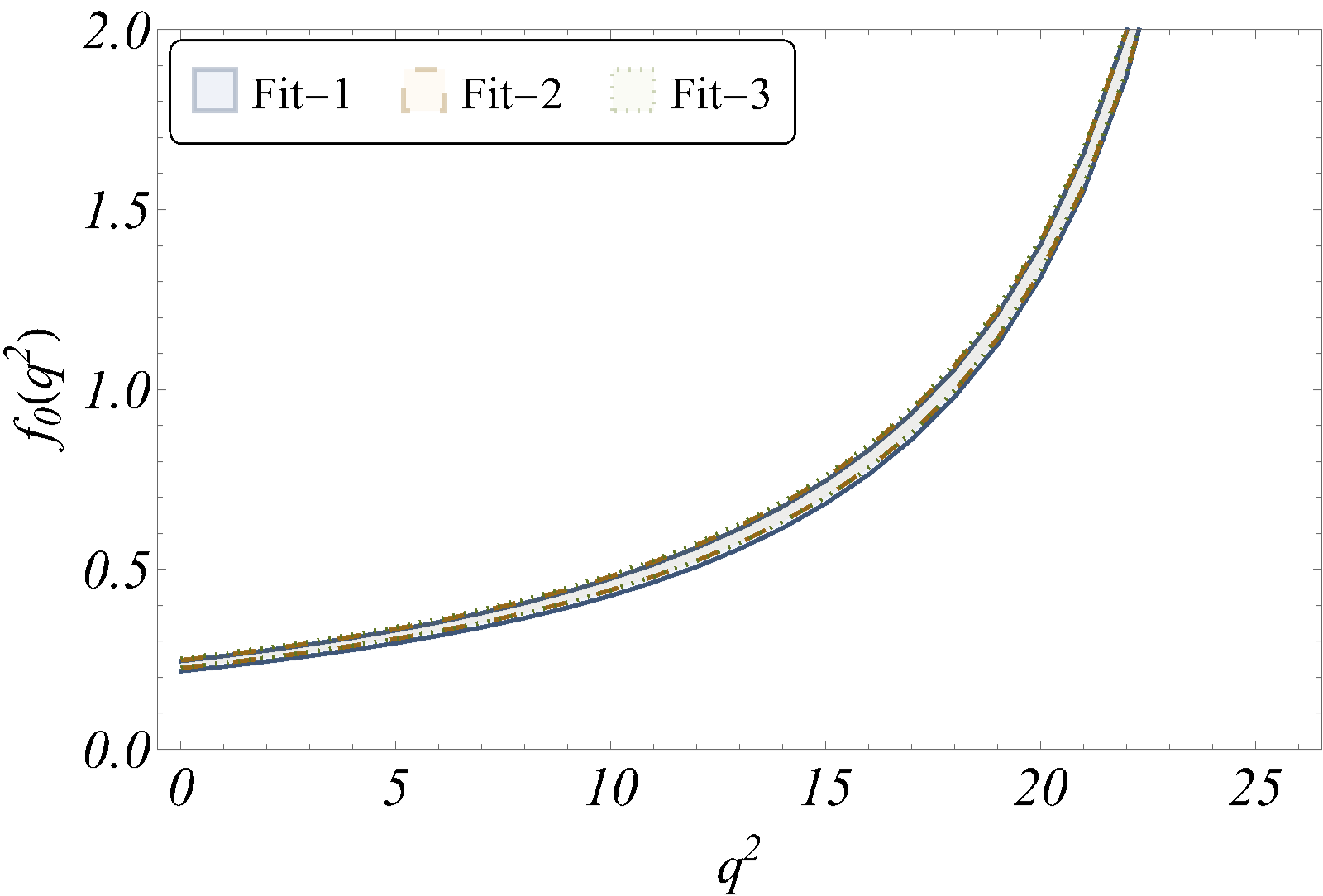}}\\
	\caption{The $q^2$-distributions of the form factors $f_{+,0,T}(q^2)$ in three different fit scenarios for which the results are presented in table \ref{tab:Btopifitresults}.}
	\label{fig:piform}
\end{figure}

\begin{table}[t]
	\begin{tabular}{|c|c|c|}\hline
		\textbf{Parameters}     & \textbf{HFLAV(2019) average \cite{HFLAV:2019otj} }  &  \textbf{`Fit-2' + BaBar 2011 data points \cite{delAmoSanchez:2010af} }\\ 
		& \textbf{ + Lattice \& LCSR} &  \\ \hline
		$\chi ^2\text{/dof}$  &   $\text{28.4296/47}$  &  $\text{102.783/96}$   \\
		$\text{p-value}$  & $0.985276$  &   $0.299333$      \\\hline
		$|V_{ub}|$ & \text{0.003833(88)} & \text{0.003782(88)} \\\hline
		$a_0^+$  &  $\text{0.2491(70)}$  &  $\text{0.2489(71)}$  \\
		$a_1^+$  &  $\text{-0.564(73)}$  &  $\text{-0.560(77)}$ \\
		$a_2^+$  &  $\text{0.25(26)}$  &  $\text{0.21(27)}$ \\
		$a_3^+$  &  $\text{0.64(20)}$  &  $\text{0.61(20)}$  \\
		$a_1^0$  &  $\text{0.548(74)}$  &  $\text{0.549(91)}$ \\
		$a_2^0$  &  $\text{1.62(28)}$  &  $\text{1.62(35)}$  \\
		$a_3^0$  &  $\text{1.80(30)}$  &  $\text{1.81(37)}$  \\
		$a_0^T$  &  $\text{0.241(11)}$  & $\text{0.252(17)}$  \\
		$a_1^T$  &  $\text{-0.63(10)}$  &  $\text{-0.60(12)}$   \\
		$a_2^T$  &  $\text{-0.060(410)}$  &  $\text{-0.043(476)}$   \\
		$a_3^T$  &  $\text{0.31(37)}$  &  $\text{0.31(43)}$  \\\hline
	\textbf{Observables}  &  \textbf{Respective predictions}  &  \textbf{Respective predictions }  \\ \hline
		$\text{R($\pi $)}$  &  $\text{0.677(10)}$  &  $\text{0.677(12)}$  \\
		$R_{\tau }^{\tau }\text{($\pi $)}$  &  $\text{0.935(40)}$  &  $\text{0.943(42)}$  \\
		$R_{\tau }^{\mu }\text{($\pi $)}$  &  $\text{0.00420(18)}$  &  $\text{0.00424(19)}$  \\
		$R_{\mu }^{\tau }\text{($\pi $)}$  &  $\text{0.633(30)}$  &  $\text{0.638(30)}$  \\\hline
	\end{tabular}
\caption{Predictions of the observables defined in eqns.~\ref{eq:Rpiobs} corresponding to the fits of the 2019 averaged $B\to\pi$ binned dataset given in HFLAV~\cite{Amhis:2019ckw} along with the inputs from lattice and LCSR (second column) and with all the data points defined in `Fit-2' plus the  BaBar 2011 data points \cite{delAmoSanchez:2010af} (third column). In  we have not considered tensorial inputs. The last four rows presents the predictions of the observables in the two fits, respectively.}
\label{tab:BtopifitresultsWBabar}
\end{table}

\subsection{$B\to\pi$}\label{subsec:Btopi}
As mentioned earlier, in order to predict the values of the observables in SM and different NP scenarios, we need to first extract the shape of the form factors $f_i(q^2)$. To get the shape, we have used Bharucha-Straub-Zwicky (BSZ) parametrization \cite{Straub:2015ica}. For BSZ, the parametrization of any form-factor reads: 
\begin{equation}
f_i(q^2) = \frac{1}{1 - q^2/m_{R,i}^2} \sum_n a_n^i \, [z(q^2)-z(0)]^n\,,
\label{eq:bszexp}
\end{equation}
where $m_{R,i}$ denotes the mass of sub-threshold resonances compatible with the quantum numbers of the respective form factors and $a_k^i$s are the coefficients of expansion. The conformal map from $q^2$ to z is given by :
\begin{equation}
z(t) = \frac{\sqrt{t_+-t}-\sqrt{t_+-t_0}}{\sqrt{t_+-t}+\sqrt{t_+-t_0}}\,,  
\end{equation}
where
$t_\pm \equiv (m_B\pm m_{\pi})^2$ and $t_0\equiv t_+(1-\sqrt{1-t_-/t_+})$. $t_0$ is a free parameter that governs the size of $z$ in the semileptonic phase space. The details are provided in \cite{Straub:2015ica}. 

As mentioned in section \ref{sec:theory}, the decay rate $\Gamma(B\to \pi \mu(e)\nu)$ requires complete knowledge of the form factors $f_{+,0}(q^2)$ while for $\Gamma(B\to \pi \tau\nu)$, we need to know $f_T(q^2)$ also. For all these form factors, we consider the expansion given in eq. \ref{eq:bszexp} up to $n=3$. In order to extract the coefficients $a_n^i$ of the $z(q^2)$ expansion of the different form factors, we carry out the following fits:
\begin{itemize}
	\item \textbf{Fit-1:} We fit the coefficients of each of the form factors $f_{+,0,T}$ to lattice and LCSR datapoints. LCSR data have been provided by and taken from~\cite{Leljak:2021vte} and~\cite{Gubernari:2018wyi}. The Lattice data has been taken from UKQCD~\cite{Flynn:2015mha} and MILC~\cite{Lattice:2015tia} for $f_{+,0}$ and from MILC~\cite{Bailey:2015nbd} for $f_T$. Details about the Lattice and LCSR data and how we use them can be found in our previous article~\cite{Biswas:2021qyq}\footnote{The LCSR data due to~\cite{Leljak:2021vte} was not used in the analysis of ref.~\cite{Biswas:2021qyq}. However, we have checked that their inclusion does not result in substantial change for the fit values and amounts to a tiny reduction in the uncertainties for the final value of $V_{ub}$.}. Note that the result of this fit is not affected by experimental data. 
	
	\item \textbf{Fit-2:} We add experimental binned data on the Branching Ratios (BR's) for $B\to\pi l\nu$ provided by BaBar~\cite{Lees:2012vv} and Belle~\cite{Ha:2010rf,Sibidanov:2013rkk} to the datalist corresponding to `Fit-1' and fit the coefficients of the form factors along with $|V_{ub}|$. We do not include the Babar 2011~\cite{delAmoSanchez:2010af} data as they do not provide the correlations among the bins for the individual charged and neutral semileptonic B decays, and including the combined data yields a poorer fit, as has been discussed in~\cite{Lattice:2015tia,Biswas:2021qyq}.
	
	\item \textbf{Fit-3:} In this scenario, we prepare synthetic data points by normalizing the branching fraction $\mathcal{B}(B\to \tau\nu)$ charged by binned branching ratios used in the previous set. In addition to these synthetic data points, we have used LCSR and lattice as discussed in `Fit-1'. In this ratio, $|V_{ub}|$ cancels. Therefore we don't need to fit it. We use $f_B=0.19\pm 0.0013$~\cite{pdgrev} as a nuisance parameter in this fit. As mentioned earlier, the $\mathcal{B}(B\to \tau\nu)$ is sensitive to NP effects. Therefore, for this fit-scenario, we have done two different analyses: (i) We fit only the parameters of the form-factors without considering NP effects in $B\to\tau\nu$ decays, (ii) We simultaneously fit the parameters of the form-factors alongside the new WCs, keeping one NP WC at a time as the free parameter\footnote{We have also carried out a fit where we consider the NP parameter to be free while all others ($f_B$ and the form factor parameters) are taken to be nuisance. We have verified that the results of these two fits are consistent with each other, especially corresponding to the leading order form factor coefficients. As a result of this, the SM prediction for the observables are also consistent within $1\sigma$ for both the fits.}. 
\end{itemize}

Note that we have considered the BSZ expansion of the form factors up to order $n=3$ given in eq.~\ref{eq:bszexp} in all these fits. As pointed out in our earlier publication \cite{Biswas:2021qyq}, we checked that the optimal description of the synthetic data from LCSR and lattice is obtained when both $f_0$ and $f_+$ are truncated at $n= 3$. We have done this by truncating the series at different orders, starting from 0 to 4 for both $f_+$ and $f_0$ and carrying out a model selection procedure incorporating Akaike Information criteria (AIC) and the modified AIC$_c$. The procedure has been followed even after the inclusion of experimental data and has led us to a similar conclusion. The details about these procedures of model selection can be found in our other articles. For example, see ref. \cite{Jaiswal:2020wer}. We have noted slight changes between the fit results for $n=2$ and $n=3$. Also, the respective predictions of the observables have small differences. We will present the rest of our results for $n=3$ since we are getting a reasonably good fit with corresponding to this scenario. In particular, the higher-order coefficients are reasonably well constrained.

\begin{table}[t]
	\small
	\begin{tabular}{|c|c|c|c|c|c|c|}
		\hline
		\multirow{3}{*}{\textbf{Observables}} & \multicolumn{6}{c|}{\textbf{Scenarios}}\\
		\cline{2-7}
		&\multicolumn{2}{c|}{$C_{V_1}$} &\multicolumn{2}{c|}{$C_{V_2}$} &\multicolumn{2}{c|}{$C_P$}  \\
		\cline{2-7}
		& $\text{Sol-1}$  &  $\text{Sol-2}$  &  $\text{Sol-1 }$  &  $\text{Sol-2 }$  &  $\text{Sol-1 }$  &  $\text{Sol-2 }$  \\
		&  -0.20(13) & -1.80(13)  & 0.20(13)   & 1.80(13)   & -0.055(36)  & -0.481(36) \\
		\hline 
		$\text{R($\pi $)}$  &  $\text{0.44(15)}$  &  $\text{0.44(15)}$  &  $\text{1.00(22)}$  &  $\text{5.41(53)}$  &  $-$  &  $-$  \\
		$R_{\tau }^{\tau }\text{($\pi $)}$  &  $\text{0.988(65)}$  &  $\text{0.988(65)}$  &  $\text{0.43(24)}$  &  $\text{0.080(20)}$  &  $\text{0.63(21)}$  &  $\text{0.63(21)}$  \\
		$R_{\tau }^{\mu }\text{($\pi $)}$  &  $\text{0.0070(23)}$  &  $\text{0.0070(23)}$  &  $\text{0.00306(71)}$  &  $\text{0.000568(66)}$  &  $\text{0.00444(29)}$  &  $\text{0.00444(29)}$  \\
		$R_{\mu }^{\tau }\text{($\pi $)}$  &  $\text{0.43(15)}$  &  $\text{0.43(15)}$  &  $\text{0.43(15)}$  &  $\text{0.43(15)}$  &  $\text{0.43(15)}$  &  $\text{0.43(15)}$  \\
		\hline
	\end{tabular}
	\caption{The results of the observables defined in eq. \ref{eq:Rpiobs} corresponding to the two NP solutions obtained from fitting the `Fit-3' dataset considering NP effects in $\mathcal{B}(B\to \tau\nu)$. Sol-1 corresponds to the solution close to the SM. Sol-2 corresponds to the solution away from the SM.}
	\label{tab:rtaumunp}
\end{table}

The results of the different fits are given in table \ref{tab:Btopifitresults}. We can fit the data well in all the scenarios. The quality of fit is reduced in `Fit-2', which includes the experimental data. In all the fits, the parameters are reasonably well constrained and the allowed values of the parameters are consistent with each other. In `Fit-2', the extracted errors of parameters are less compared to the other two fit scenarios.  The $q^2$-distributions of the form factors $f_+(q^2)$, $f_0(q^2)$ and $f_T(q^2)$ in three different fit scenarios have been shown in figure \ref{fig:piform}. Note that throughout the kinematically allowed $q^2$ region, the form factors are consistent in all the three fit scenarios. In the appendix in table \ref{tab:appffq2} we have presented the best fit values along with respective errors of the form factors for a few values of $q^2$. The predictions can be compared in the different fit scenarios, which agree with each other. In table~\ref{tab:SMpredrho} (last four rows), we have provided the SM predictions of the observables discussed in section \ref{sec:theory}. Note that within the error bars, all the predictions are consistent with each other. In `Fit-1', all the predictions have large errors compared to that obtained in `Fit-2' and `Fit-3', which is not surprising since the LCSR inputs have large errors. We are using experimental inputs in `Fit-2' and `Fit-3', and the predictions are extremely consistent with each other. In `Fit-3', we have used synthetic data for $R_{\tau}^\mu(\pi)$, and these data points are generated using the experimental input on $\mathcal{B}(B\to \tau\nu)$ which is sensitive to NP effects. However, according to our assumption, the data points used in `Fit-2' are not sensitive to any NP effects. The consistency between the results of different fits indicates that at the moment, we don't see the possibility/requirement of large NP effects in $b\to u\tau\nu$ decays though it is allowed considering the error in the SM predictions and a few specific measurements.  

For completeness, we have analyzed the Babar 2011 data~\cite{delAmoSanchez:2010af} alongside the data points in the scenario `Fit-2'. We have also carried out a separate fit using the heavy flavour averaging group (HFLAV) average of all the available experimental data on $B\to \pi \ell\nu$ decays given in ref. \cite{HFLAV:2019otj}, together with the respective lattice and LCSR information as additional inputs. The fit results and the respective predictions of the associated observables are shown in table \ref{tab:BtopifitresultsWBabar}. The predictions in both the fits are extremely consistent with each other. The results obtained can be compared with the respective predictions in `Fit-2'. The best fit points have shifted a bit, but are consistent with each other within their respective 1-$\sigma$ error bars.

\begin{figure}[t]
	\centering
	\subfloat[]{\includegraphics[width=0.40\textwidth]{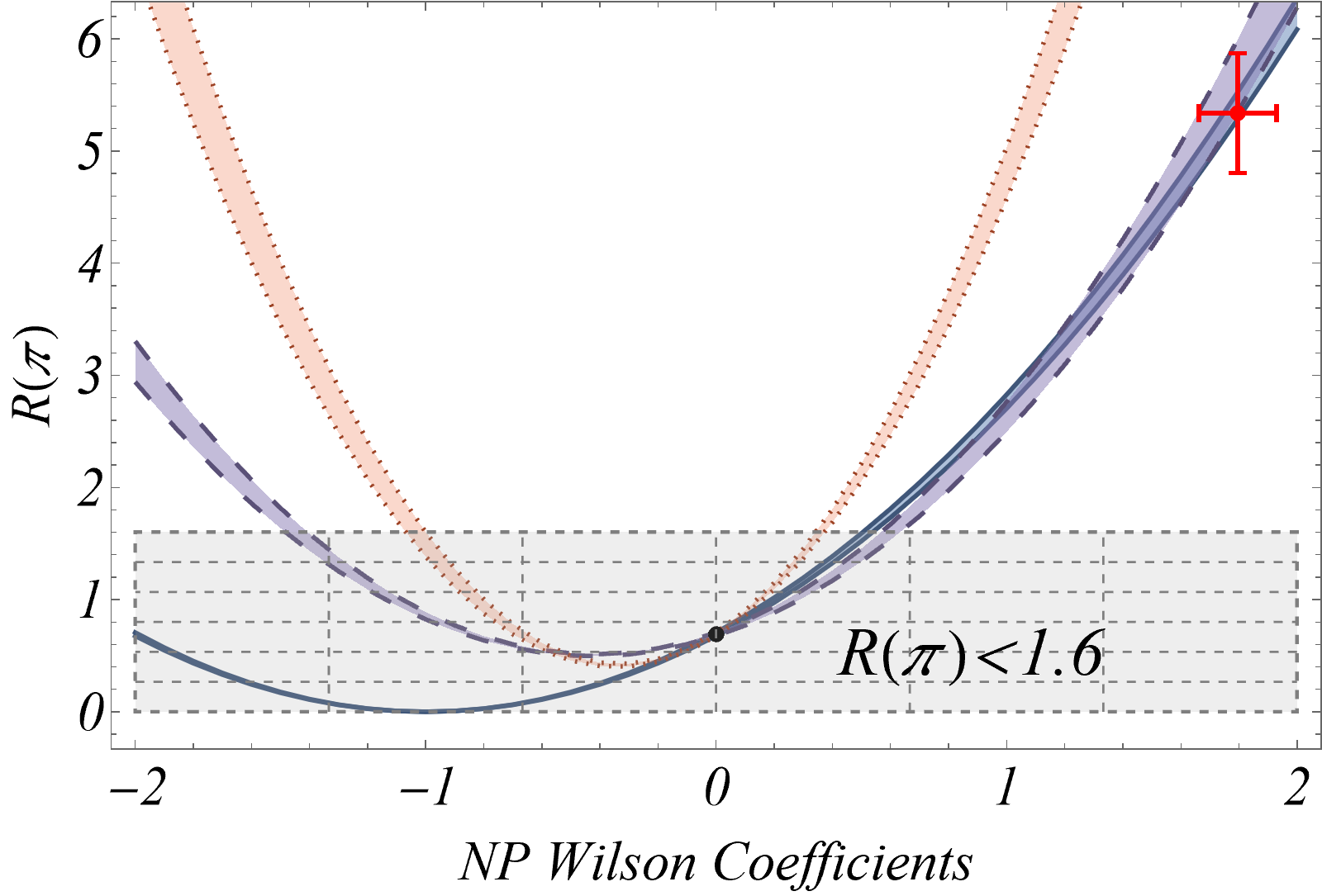}\label{fig:Rpi}}~
	\subfloat[]{\includegraphics[width=0.40\textwidth]{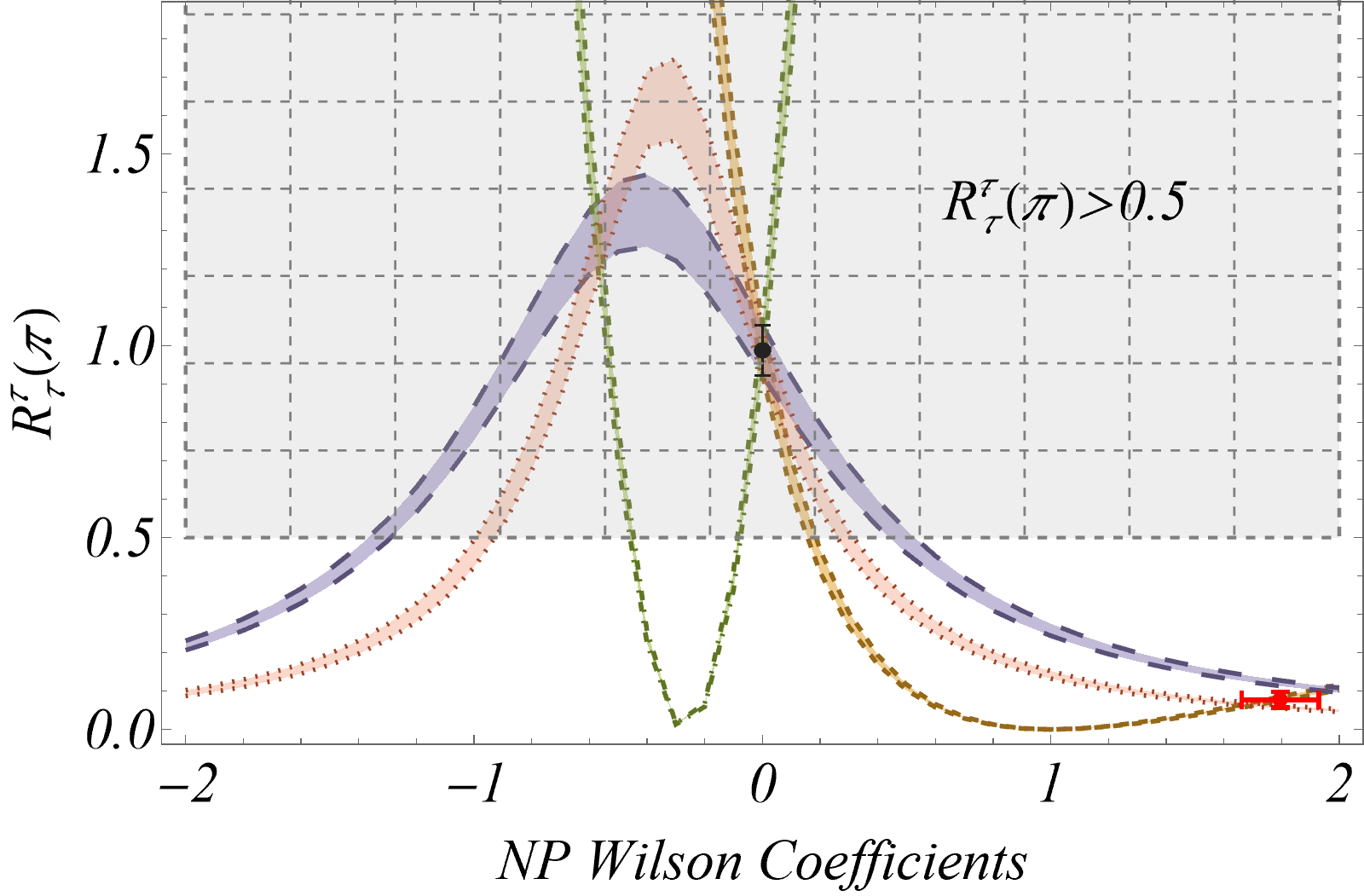}\label{fig:Rtautau}}\\
	\subfloat[]{\includegraphics[width=0.40\textwidth]{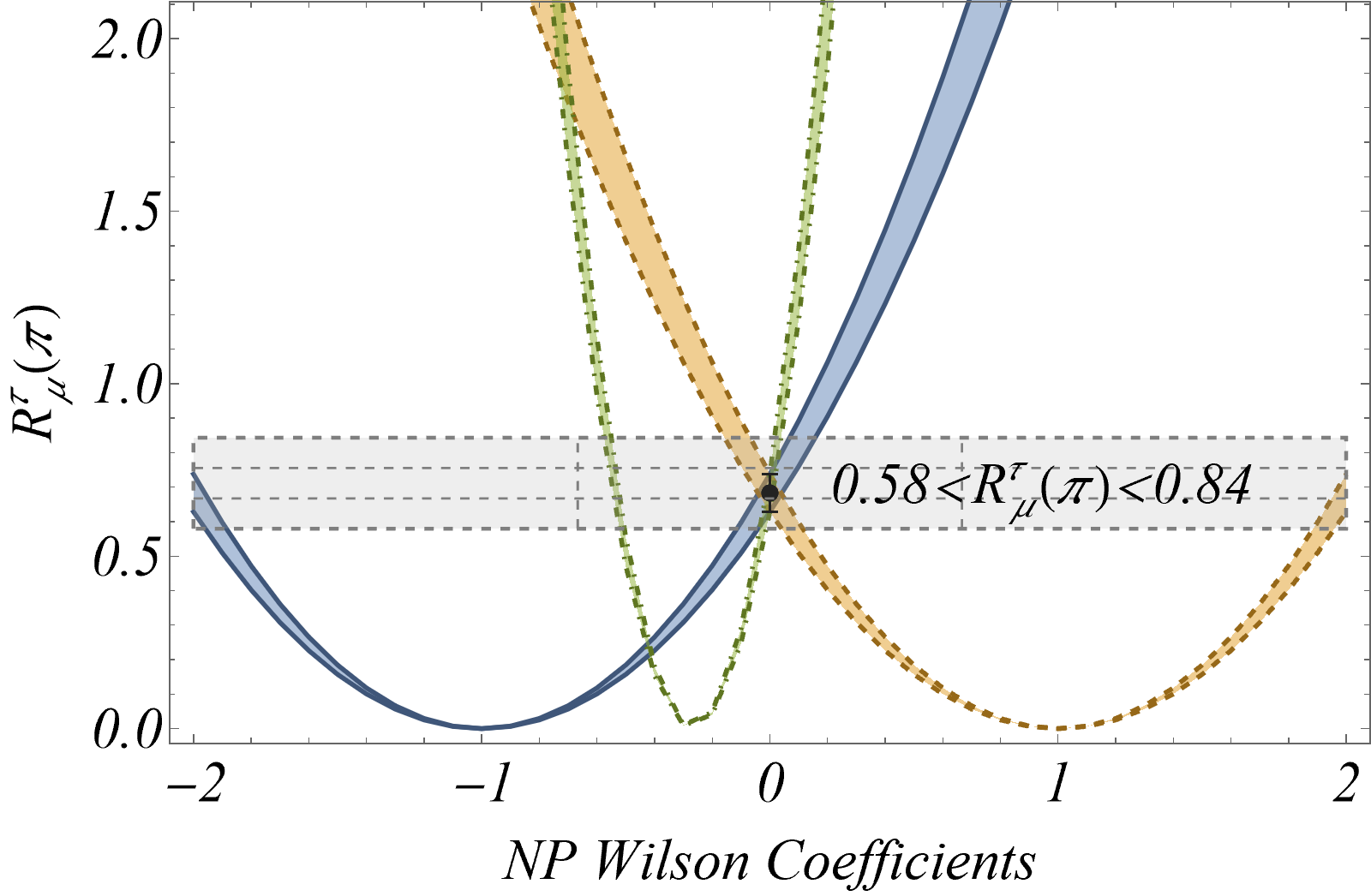}\label{fig:Rtaumu}}~~~~~
	\subfloat[]{\includegraphics[width=0.08\textwidth]{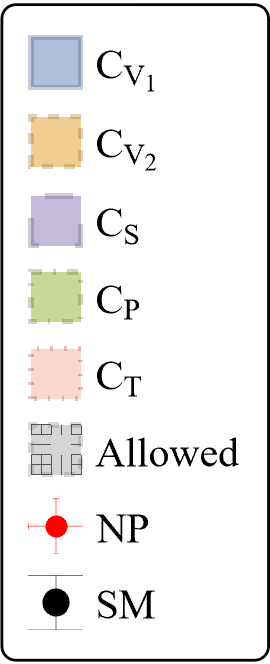}}\\
	\caption{Variations of the observables $R(\pi)$, $R_{\tau }^{\tau }(\pi)$ and $R_{\tau }^{\mu }(\pi)$ w.r.t. new WC's (real), taken one at a time. The grey hatched regions are allowed by the current experimental limits. The exclusion region is derived from the Belle estimate for the upper limit corresponding to $BR(B^0\to\pi^-\tau^+\nu_{\tau})<2.5\times 10^{-4}$ at $90\%$ confidence level. The values of the observables corresponding to out NP solutions from fit have also been shown. The corresponding best fit values for the NP WC's have been provided in the legend.}
	\label{fig:obsvarnp}
\end{figure}

As mentioned eralier, in `Fit-3' we have also carried out an analysis considering NP effects in $B\to\tau\nu$ decays. This decay mode is sensitive to $C_{V_1}$ or $C_{V_2}$ and $C_P$, but not to $C_T$ and $C_S$. The fit results for the new WCs along with the respective errors are shown in table \ref{tab:rtaumunp} as `Sol-1' and `Sol-2', respectively. Note that the solutions `Sol-2' are largely deviated from zero (SM) while `Sol-1' is closer to (but inconsistent with) zero. We have carried out the fitting procedure considering one WC at a time, though it is possible to build models where new physics may contribute simultaneously to all the WCs. However, it is difficult to fit all of them simultaneously in a frequentist $\chi^2$-minimization procedure since the decay rate $\Gamma(B\to\tau\nu)$ as given in eq. \ref{eq:brlep}. is sensitive to $\left|1+C_{V_1}-C_{V_2}+\frac{m_B^2}{m_\tau(m_b+m_u)}C_P\right|^2$. Different possible solutions of the combination $[C_{V_1},C_{V_2},C_P]$ will correspond the same value of $\chi^2_{min}$ and hence we won't get a stable solution. We can fit only one new WC at a time, with the solutions roughly following the relations: $C_{V_2} = -C_{V_1} \approx \frac{m_B^2}{m_\tau(m_b+m_u)}C_P$ as can be seen from table \ref{tab:rtaumunp}. At present, it will be helpful to keep the analysis as simple as possible since that will help us understand the pattern of new physics effects in the associated observables with much certainty. To constrain the more complex NP scenario(s) we need data on $B\to\pi \tau\nu$ decays.

Using the results for the new WCs, we have predicted $R(\pi)$, $R_{\tau }^{\tau }(\pi)$, and $R_{\tau }^{\mu }(\pi)$ in the respective NP scenarios. In the predictions of $R(\pi)$ and $R_{\tau }^{\tau }(\pi)$, the form factors are obtained from the fit results of `Fit-1' while that for $R_{\tau }^{\mu }(\pi)$ we have used the fit results of `Fit-3' considering NP effects in $B\to \tau\nu$ decays. In figure \ref{fig:obsvarnp}, we have shown the NP sensitivities of different observables as mentioned above along with the regions allowed by the current experimental limits given in eq. \ref{eq:expbounds}. For the purpose of illustration, we have provided predictions of these observables for a few benchmark values of the new WCs in table \ref{tab:nppipred} in the appendix. Some observations from table \ref{tab:rtaumunp} and figure \ref{fig:obsvarnp}:
\begin{itemize}  
		\item The current data on $R_{\mu }^{\tau }(\pi)$ prefers negative solutions for real $C_{V_1}$ and positive solutions for $C_{V_2}$. 
	
	\item $R(\pi)$ is not sensitive to $C_P$. In the allowed regions of $C_{V_1}$ (and with an educated guess that the size of the NP will be small if present), the value of $R(\pi)$ will reduce from the respective SM value. In contrast, the allowed value of $R(\pi)$ for $C_{V_2}$ will increase from that of its SM prediction. These observations will be helpful to distinguish the effects of the operators $\mathcal{O}_{V_1}$ and $\mathcal{O}_{V_2}$ from a measurement of $R({\pi})$.   
	
	\item Both the solutions of $C_{V_1}$ obtained from the fit to $R_{\tau }^{\mu }(\pi)$ are also allowed by the current experimental limit on $R(\pi)$ (fig. \ref{fig:Rpi}) while $R_{\tau }^{\tau }(\pi)$ (fig. \ref{fig:Rtautau}) is not sensitive to $C_{V_1}$.  
	
	\item One of the two solutions obtained for $C_{V_2}$ from the fit are not allowed by the current experimental limits on $R(\pi)$ and $R_{\tau }^{\tau }(\pi)$. The solution close to zero is allowed by the data. Also, a large negative value of $C_{V_2}$ is not allowed by the current data on $R_{\tau }^{\mu }(\pi)$ (fig. \ref{fig:Rtaumu}).
	
	\item As can be seen from figs. \ref{fig:Rtautau} and \ref{fig:Rtaumu} both the negative solutions for $C_P$ are also allowed by the data, however, the regions are very restricted. 
	   
	 \item  The allowed values of$C_S$, and the dependences of $R(\pi)$ and $R_{\tau }^{\tau }(\pi)$ on this WC can be seen from figures \ref{fig:Rpi} and \ref{fig:Rtautau}, respectively. Magnitudes of order one are allowed by the current data for $C_S$, and negative solutions are preferred over the positive ones.  
		 
\end{itemize}   
In table \ref{tab:NPsenspi}, the observables and their respective NP sensitivities are summarised. Once the precise measurements of all these observables are available, a comparative study of the observed deviations will pinpoint the types of new physics affecting them. However, we would like to mention that it will be hard to distinguish the contribution from $\mathcal{O}_S$ with that of $\mathcal{O}_T$.
       
\begin{table}
	\begin{tabular}{|c|c|c|c|c|c|}
		\hline
		Observables & 	\multicolumn{5}{|c|}{NP scenario}\\
		\cline{2-6}
		            & $C_{V_1}$  & $C_{V_2}$  & $C_{S}$  & $C_{P}$  & $C_{T}$ \\
		            \hline
		    $R(\pi)$  & Yes & Yes & Yes & No & Yes \\
		    $R_{\tau }^{\tau }(\pi)$& No  & Yes &Yes & Yes & Yes \\
		    $R_{\mu }^{\tau }(\pi)$ & Yes & Yes  & No & Yes  & No \\	
		    \hline 	    
	\end{tabular}
\caption{The observables along with the respective new physics scenarios which are effecting them.}
\label{tab:NPsenspi}
\end{table}

\subsection{$B \to \rho \ell \nu$ decays}\label{subsec:Btorho}

\begin{table}[t]
	\begin{tabular}{|c c|c c|c c|c c|}\hline
		\multicolumn{4}{|c|}{$\textbf{Without Babar}$}&\multicolumn{4}{|c|}{$\textbf{With Babar}$}  \\
		\hline
		\multicolumn{2}{|c|}{n =3\ \ ($p$-value 99.99\%)} & \multicolumn{2}{c|}{n = 2\ \  ($p$-value 99.89\%)}&\multicolumn{2}{|c|}{n =3\ \ ($p$-value 96.00\%)} & \multicolumn{2}{c|}{n = 2\ \  ($p$-value 93.35\%)}  \\
		\hline
		Parameters & fit values  & Parameters & fit values& Parameters &  fit values& Parameters & fit values  \\\hline
		$a_0^{A_0}$  &  $\text{0.309(17)}$   &$a_0^{A_0}$ & $\text{0.316(16)}$&$a_0^{A_0}$  &  $\text{0.290(16)}$    &$a_0^{A_0}$ & $\text{0.295(15)}$ \\
		$a_1^{A_0}$  &  $\text{-0.86(16)}$  &$a_1^{A_0}$&$\text{-0.95(13)}$&$a_1^{A_0}$  &  $\text{-0.83(16)}$   &$a_1^{A_0}$&$\text{-0.951(13)}$ \\
		$a_2^{A_0}$  &  $\text{1.5(10)}$  &$a_0^{A_1}$&$\text{0.249(13)}$&$a_2^{A_0}$  &  $\text{1.33(100)}$      &$a_0^{A_1}$&$\text{0.231(11)}$ \\
		$a_0^{A_1}$  &  $\text{0.247(16)}$   &$a_1^{A_1}$&$\text{0.424(50)}$&$a_0^{A_1}$  &  $\text{0.226(15)}$   &$a_1^{A_1}$&$\text{0.434(49)}$\\
		$a_1^{A_1}$  &  $\text{0.435(95)}$   &$a_0^{V}$&$\text{0.319(17)}$& $a_1^{A_1}$  &  $\text{0.450(94)}$    &$a_0^{V}$&$\text{0.298(15)}$\\
		$a_2^{A_1}$  &  $\text{0.33(30)}$  &$a_1^{V}$&$\text{-0.795(54)}$& $a_2^{A_1}$  &  $\text{0.30(30)}$  &$a_1^{V}$&$\text{-0.778(53)}$ \\
		$a_1^{A_{2}}$  &  $\text{-0.45(11)}$   &$a_1^{A_{2}}$&$\text{-0.468(51)}$&$a_1^{A_{2}}$  &  $\text{-0.36(10)}$ &    $a_1^{A_{2}}$&$\text{-0.408(47)}$ \\
		$a_2^{A_{2}}$  &  $\text{0.92(114)}$ &&&   $a_2^{A_{2}}$  &  $\text{0.45(109)}$ & &\\
		$a_0^V$  &  $\text{0.310(20)}$&&& $a_0^V$  &  $\text{0.287(19)}$   & &\\
		$a_1^V$  &  $\text{-0.79(13)}$&&& $a_1^V$  &  $\text{-0.78(13)}$   &  &\\
		$a_2^V$  &  $\text{1.84(90)}$&&& $a_2^V$  &  $\text{1.77(89)}$ &   &\\
		\hline
	\end{tabular}
	\caption{Fit results for the coefficients of the form factors $a_n^i$ for the vector ($V$), axial-vector ($A$'s) and tensor ($T$'s) form factors that contribute to semileptonic $B\to\rho$ transitions using a BSZ parametrization as mentioned in the text with and without the inclusion of the Babar data from ref.~\cite{delAmoSanchez:2010af}. We use the results of the `Fit-1' dataset (table~\ref{tab:Btopifitresults}) as nuisance for the integrated $B\to\pi$ decays used to normalize the binned $B\to\rho$ data. For $B\to\rho$, we use $n=3$ and $n=2$ parametrization while for $B\to\pi$ we use $n=4$. Note that the $a_n^{T_{(i)}}$'s do not take part in semileptonic charged current $B\to\rho$ transitions within the SM and hence are simply constrained by the LCSR datapoints.}
	\label{tab:fitffrho}
\end{table}

\begin{figure}[htbp]
	\centering
	\subfloat[]{\includegraphics[width=0.40\textwidth]{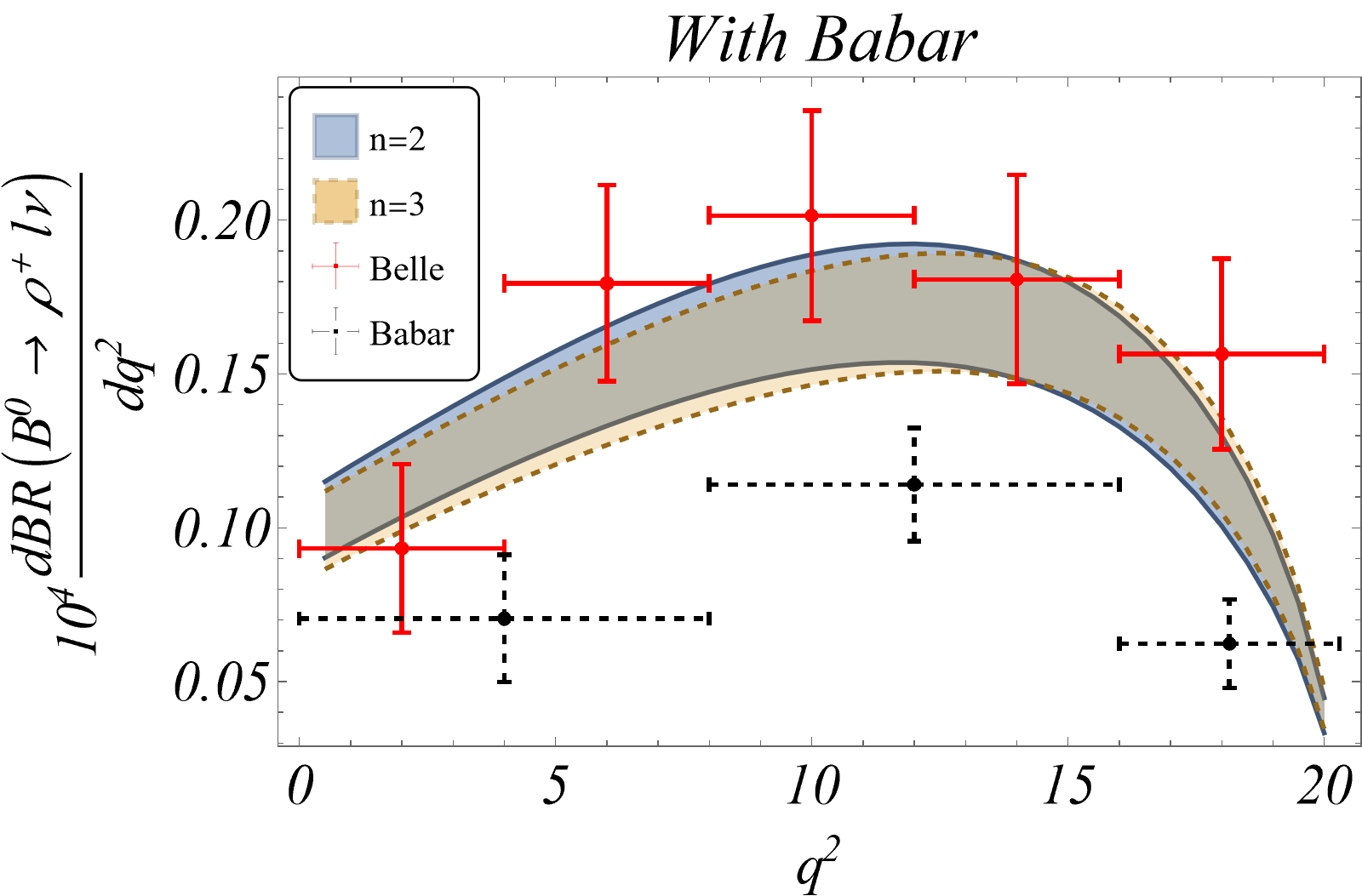}\label{fig:diffB0wBa}}~
	\subfloat[]{\includegraphics[width=0.40\textwidth]{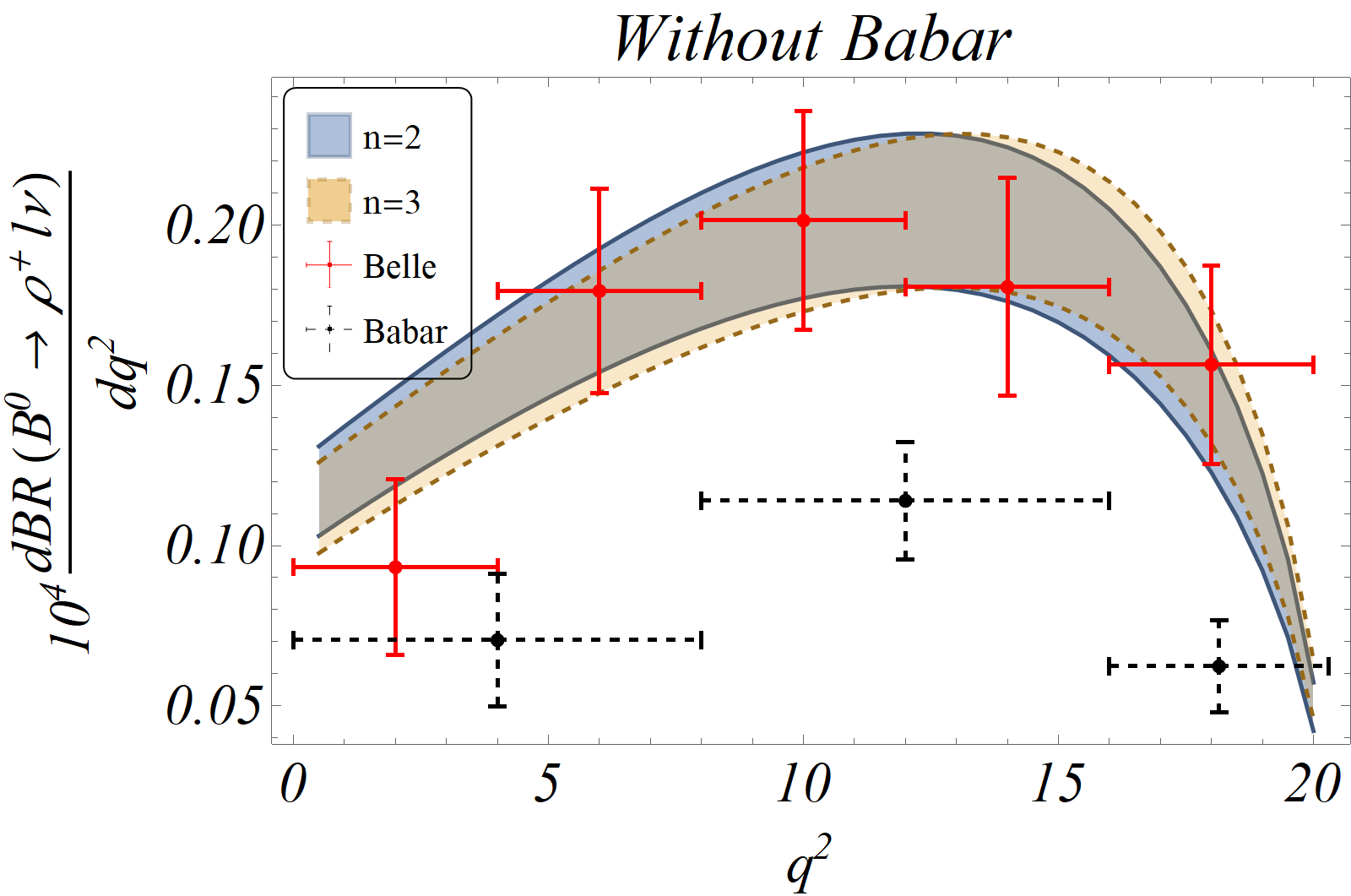}\label{fig:diffB0woBa}}\\
	\subfloat[]{\includegraphics[width=0.40\textwidth]{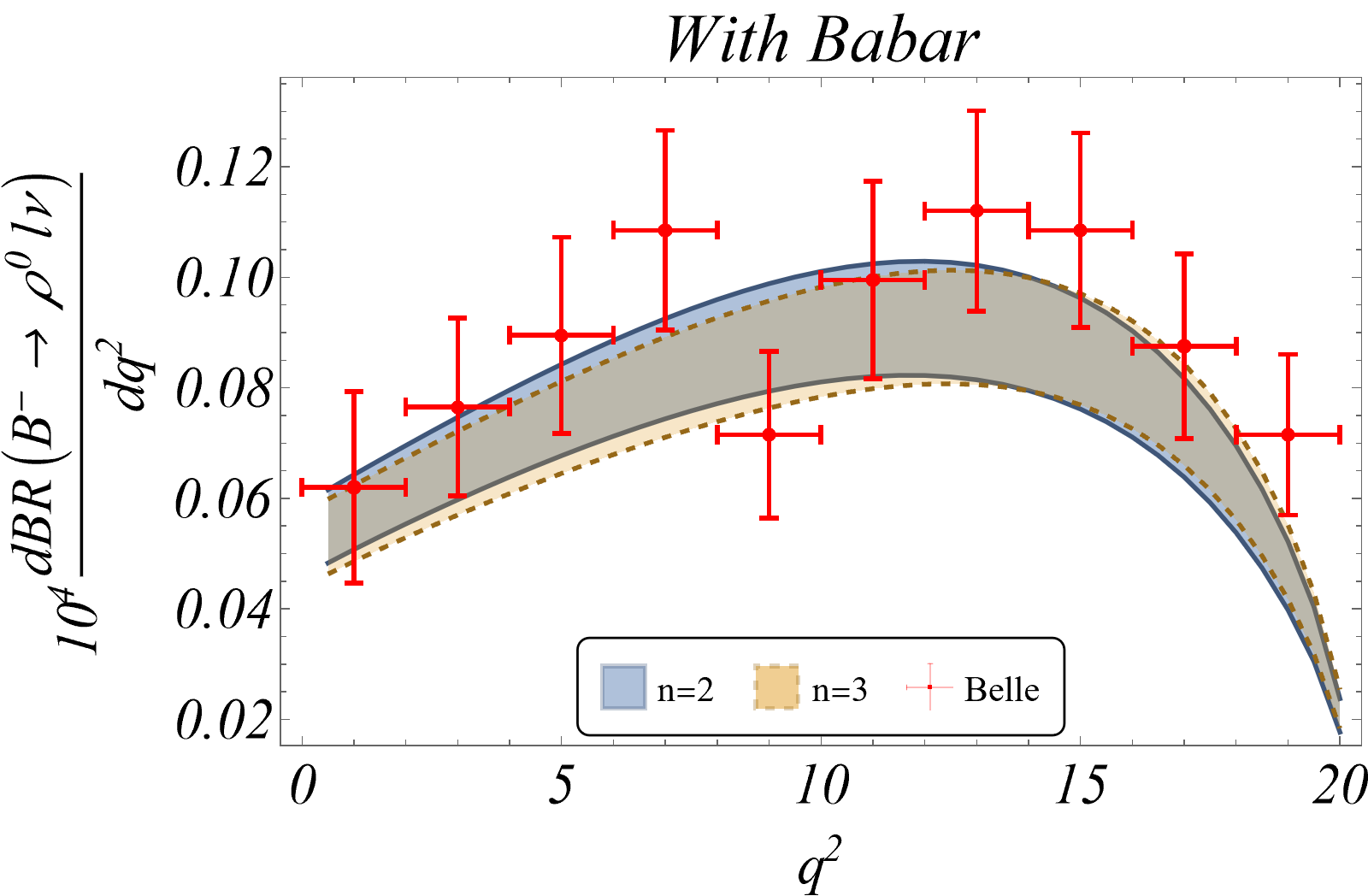}\label{fig:diffBMwBa}}~~~~~
	\subfloat[]{\includegraphics[width=0.40\textwidth]{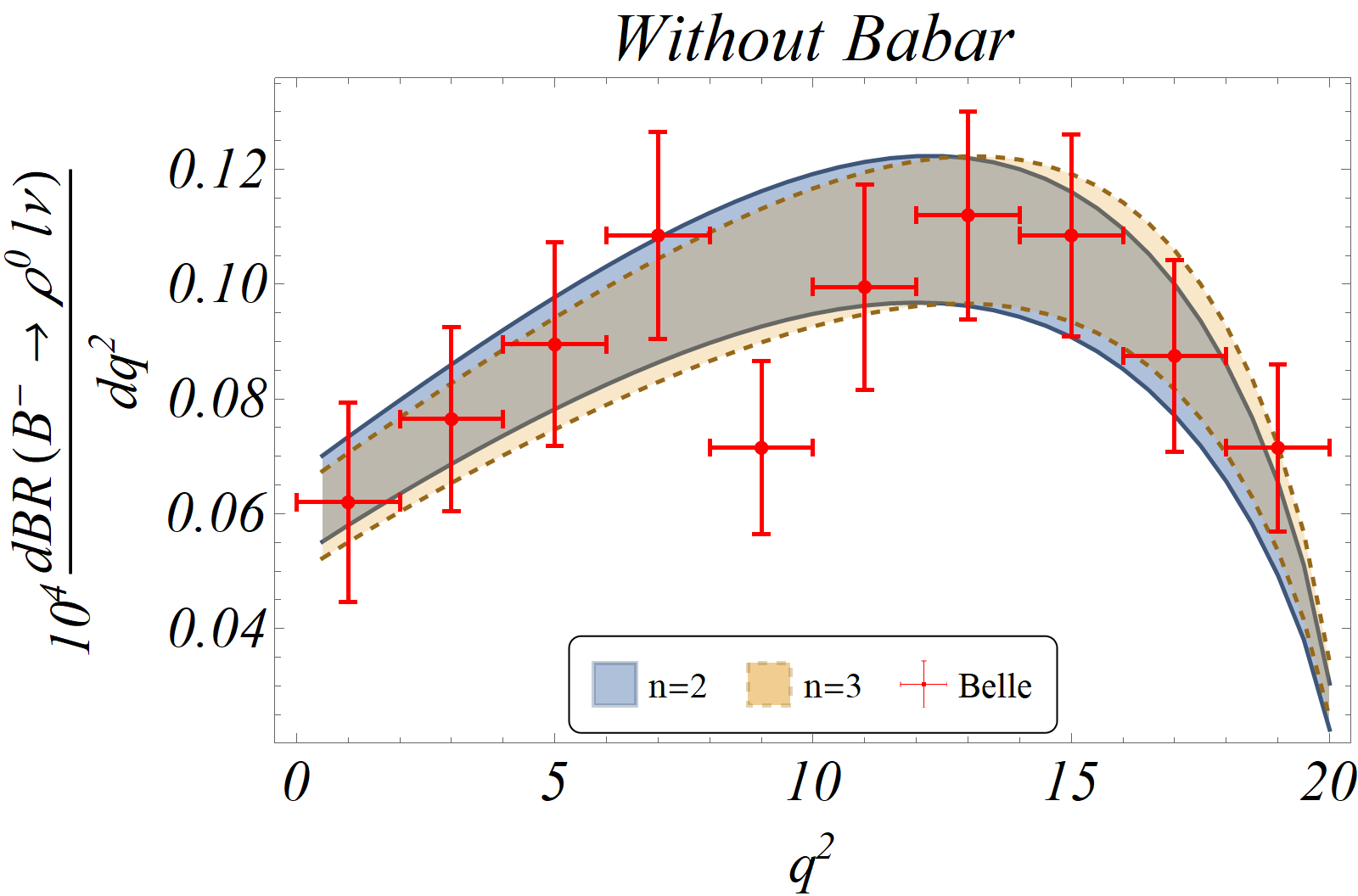}\label{fig:diffBMwoBa}}\\
	\caption{Plots showing the variation of $\frac{d\mathcal{B}}{dq^2}$ for neutral and charged $B$ to $\rho$ semileptonic decays corresponding to the fit results where the dataset includes ((a) and (b)) or does not include ((c) and (d)) the three Babar datapoints. It can clearly be seen that the inclusion of the Babar data results in the fit not being able to incorporate most of the data from Belle, whereas excluding them immediately results in a much more acceptable scenario. We have used $V_{ub}=(3.91 \pm 0.13)\times 10^{-3}$ which is an update on the result reported by~\cite{Biswas:2021qyq} after the inclusion of the LCSR data on $B\to\pi$ form factors from~\cite{Leljak:2021vte}}.
	\label{fig:diffraterho}
\end{figure}

We carry out a fit for $B\to\rho$ form factors as discussed in section \ref{sec:theory} incorporating currently available state of the art LCSR inputs alongwith binned experimental information available for $B\to\rho l\nu$ decays. As mentioned earlier, the LCSR inputs are taken from the refs.~\cite{Straub:2015ica,Gubernari:2018wyi}. On the experimental side, we have presented our main results using the binned $B\to\rho l\nu$ experimental data due to Belle~\cite{Sibidanov:2013rkk} (for $B^0$ and $B^-$ initial sates). For completeness, we have also carried out a fit including Babar data on $B \to \rho$ mode from \cite{delAmoSanchez:2010af} where the results are obtained from a combined analysis of the following four modes: $B^{0(+)}\to \pi^{+(0)}\ell\nu$ and $B^{0(+)}\to \rho^{+(0)}\ell\nu$. In order to cancel $|V_{ub}|$ out, we normalize the binned $B\to\rho l\nu$ branching fractions by the integrated branching fractions $\mathcal{B}(B\to\pi l\nu)$ of the same analysis, and the normalization is done in a way so that charged (neutral) $B\to\rho\ell\nu$ decays are normalized by charged (neutral) $B\to\pi\ell\nu$ integrated branching fractions. For the charged and neutral $B\to \pi$ modes, we have used the following data from Belle and Babar:
\begin{align}
	\mathcal{B}(\bar{B^0}\to\pi^+ l\nu)_{Belle} &= (1.49 \pm 0.09\pm 0.07)\times 10^{-4}, \ \ \  \mathcal{B}(B^-\to\pi^0 l\nu)_{Belle} = (0.80 \pm 0.08\pm
	0.04)\times 10^{-4} \text{\cite{Sibidanov:2013rkk}} \nn \\
	&  \mathcal{B}(\bar{B^0}\to\pi^+ l\nu)_{Babar} = (1.41 \pm 0.05\pm 0.07)\times 10^{-4} ~\text{\cite{delAmoSanchez:2010af}}.
	\label{eq:norinputs}
\end{align}

We normalize the combined data due to Babar by the integrated neutral $\mathcal{B}(\bar{B^0}\to\pi^+ l\nu)_{Babar}$ as given in \ref{eq:norinputs}. Both the fits with and without the Babar data points are allowed with reasonably good fit probabilities.   
The fitted coefficients of expansion of the form factor (eq.~\ref{eq:bszexp}) are given in table \ref{tab:fitffrho}. Note that the best fit points shift slightly in both the fits though they are very consistent within the given error bars. Also, we have done the analysis considering the coefficients of expansion up to order $n=3$ and $n=2$, respectively, and compared them. Note that for the fit with $n=3$, the higher-order coefficients (for example $a_2^{f_i}$) are not well constrained and have large errors. We need more precise data to constrain the higher order coefficients. The decay rates to the light leptons ($\mu$ or $e$) are insensitive to tensor form factors in the SM. Hence we are unable to constrain $T_i(q^2)$ ($i=1,2$ and $3$) using experimental data. Constraining them from LCSR alone is the best that we can do at present.

\begin{figure}[t]
	\centering
	\subfloat[]{\includegraphics[width=0.3\textwidth]{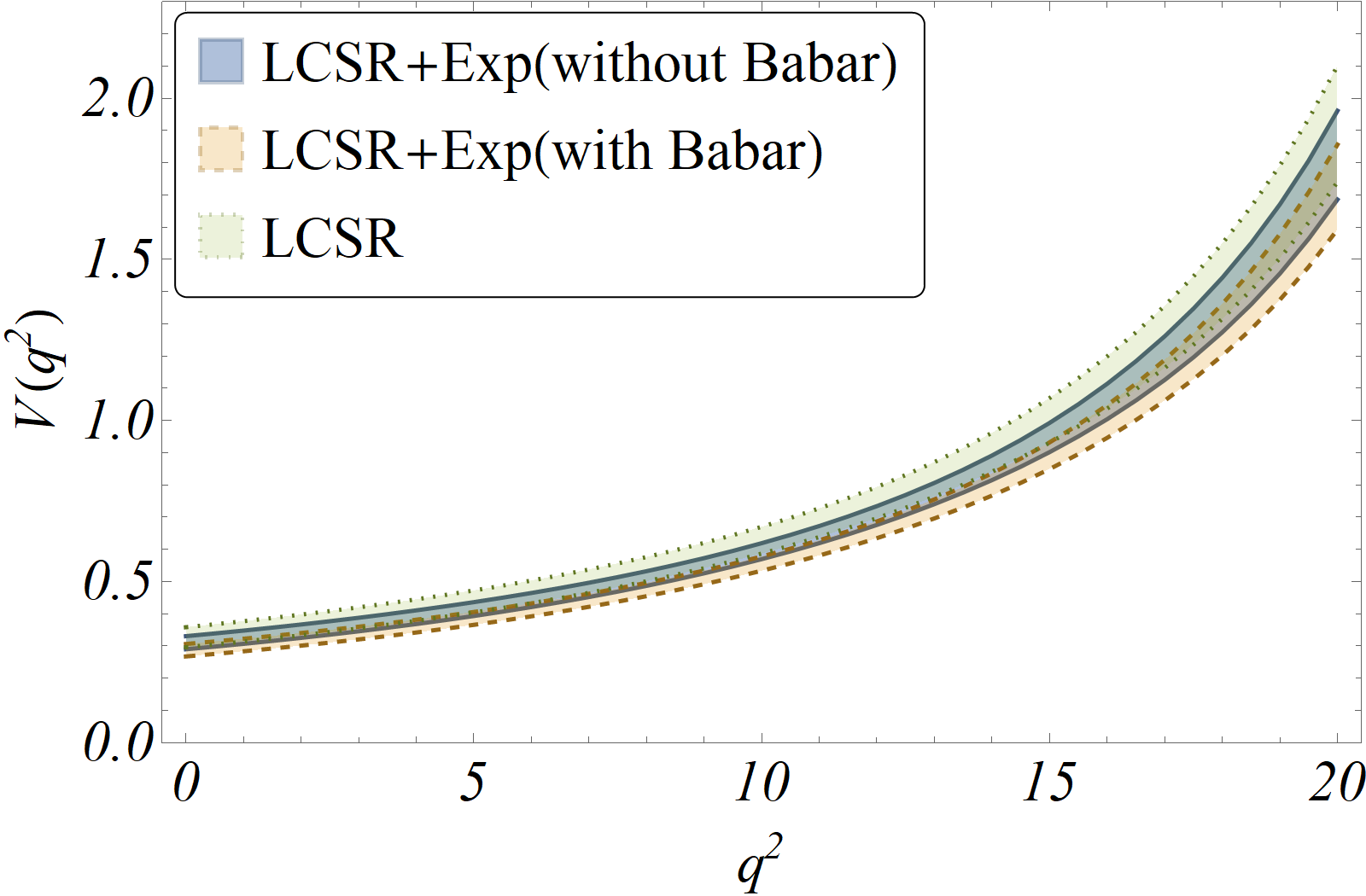}\label{fig:V0}}~~
	\subfloat[]{\includegraphics[width=0.3\textwidth]{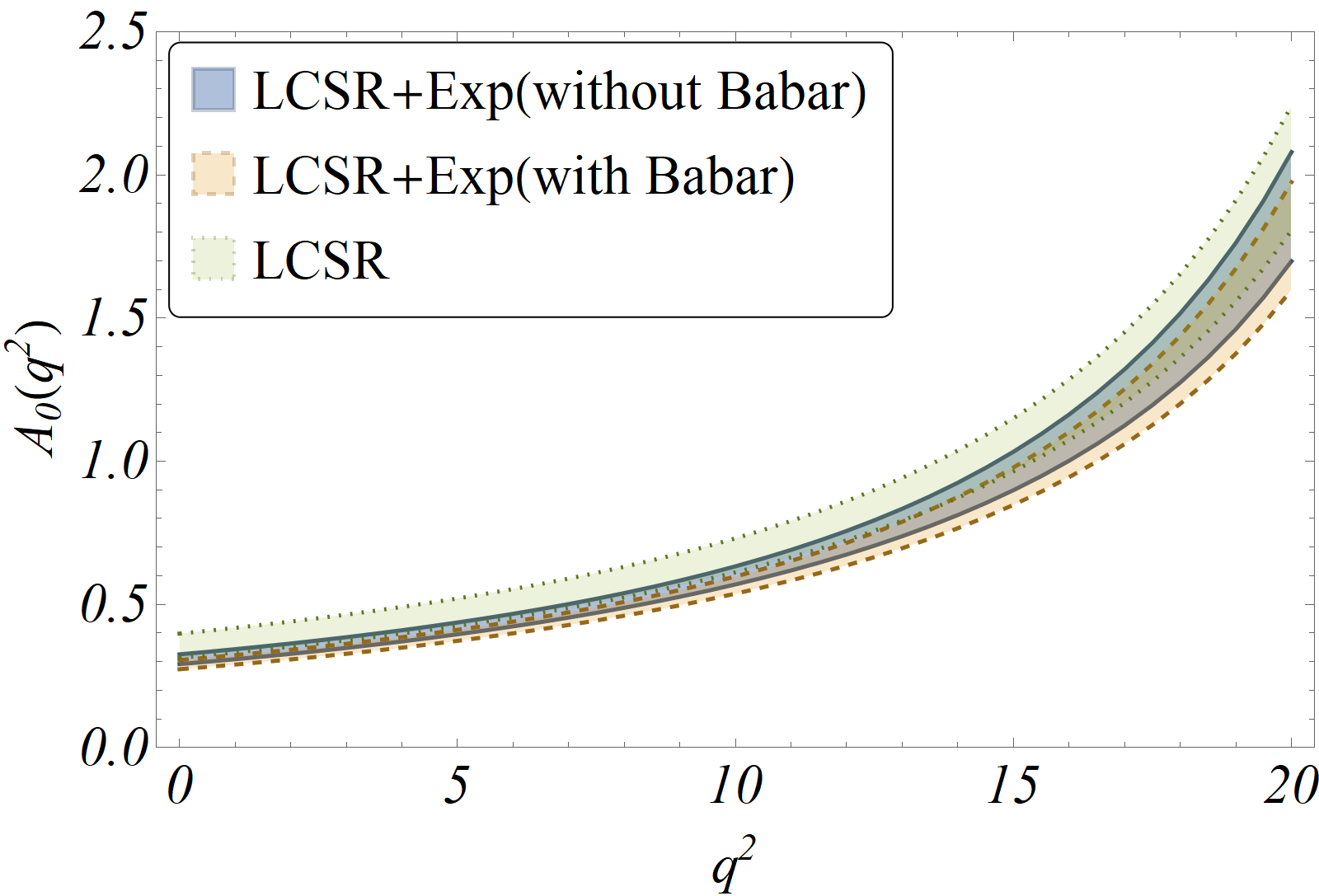}\label{fig:A0}}~~
	\subfloat[]{\includegraphics[width=0.3\textwidth]{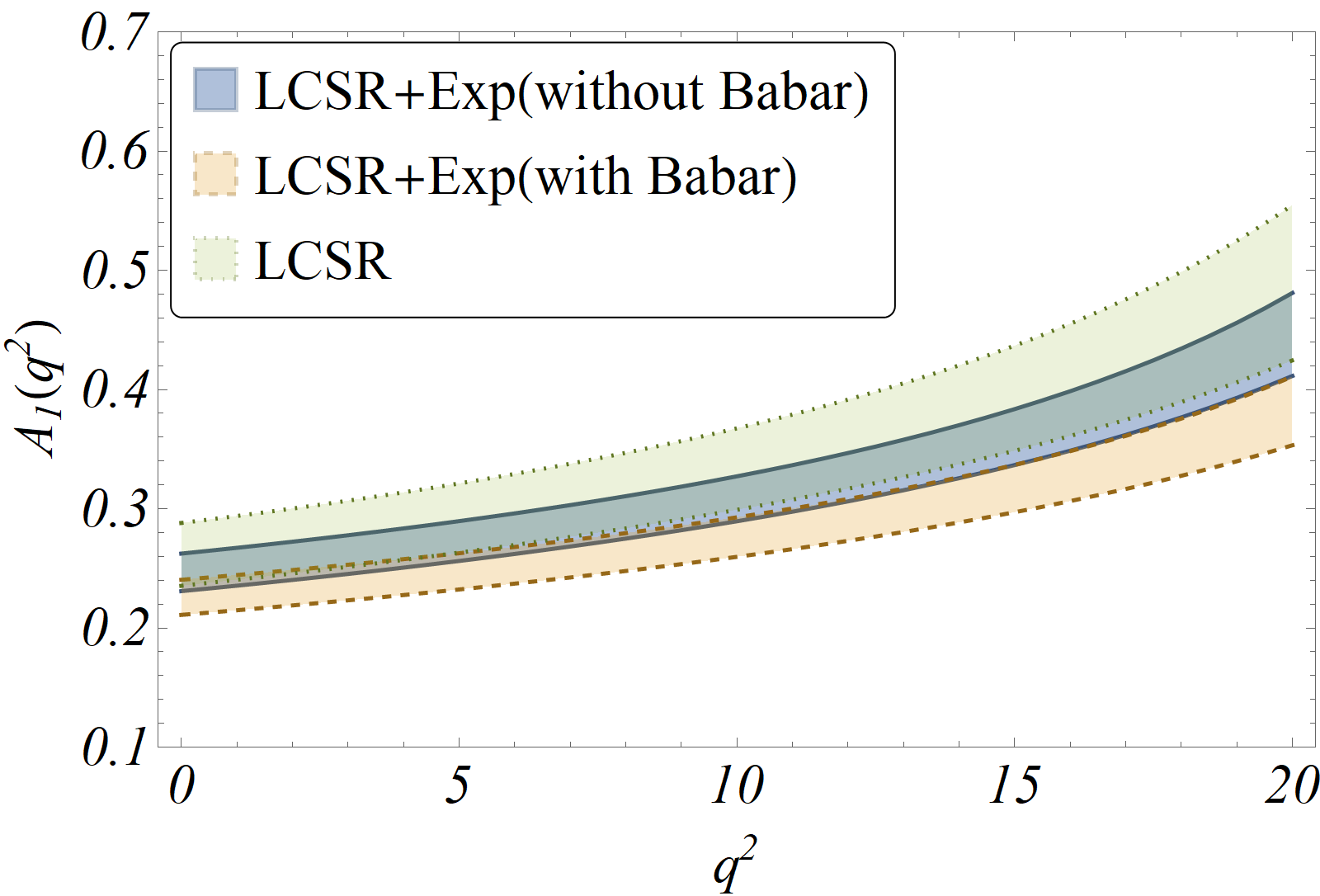}\label{fig:A1}}~~\\
	\subfloat[]{\includegraphics[width=0.3\textwidth]{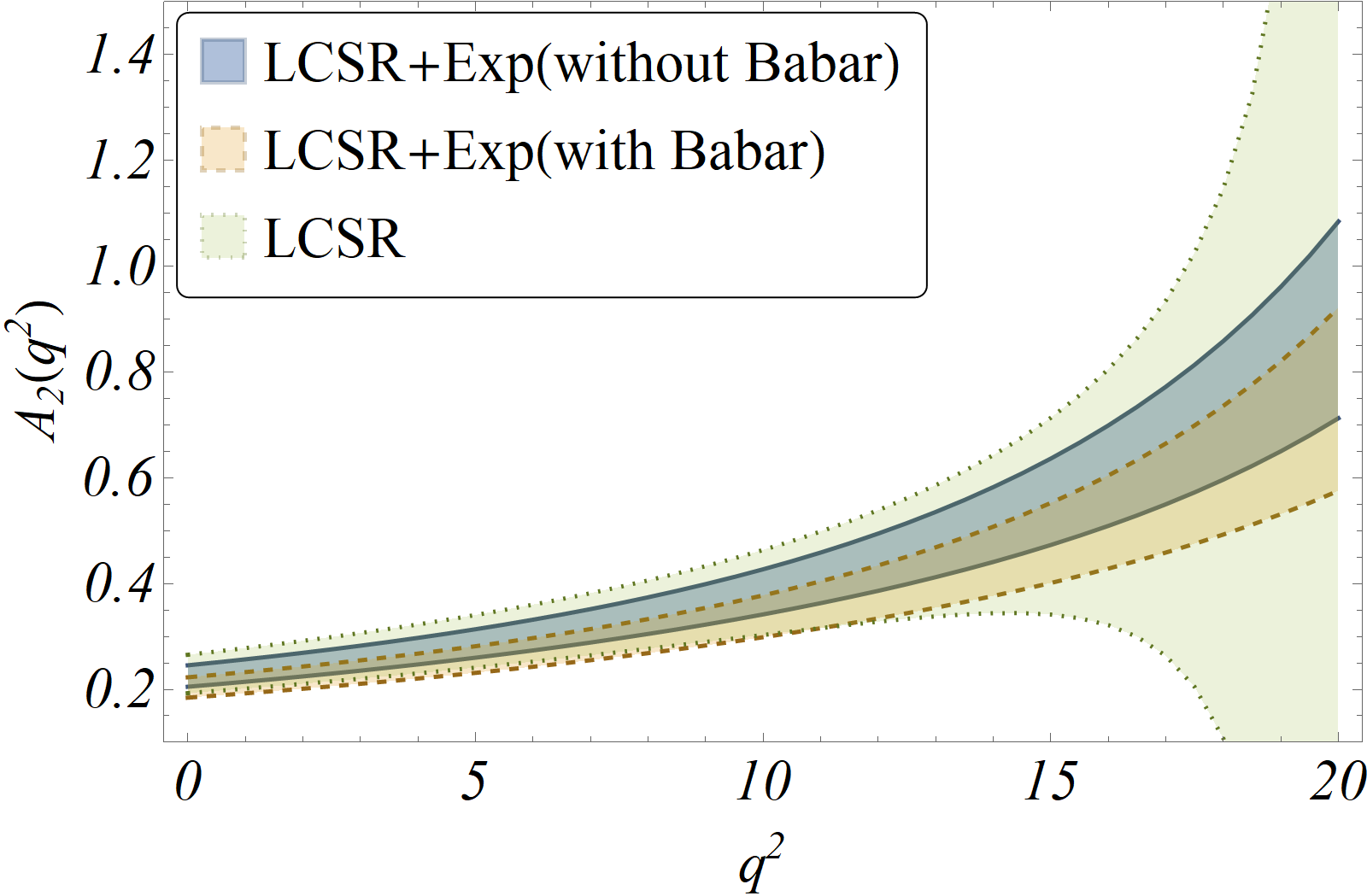}\label{fig:A2}}~~
	\subfloat[]{\includegraphics[width=0.3\textwidth]{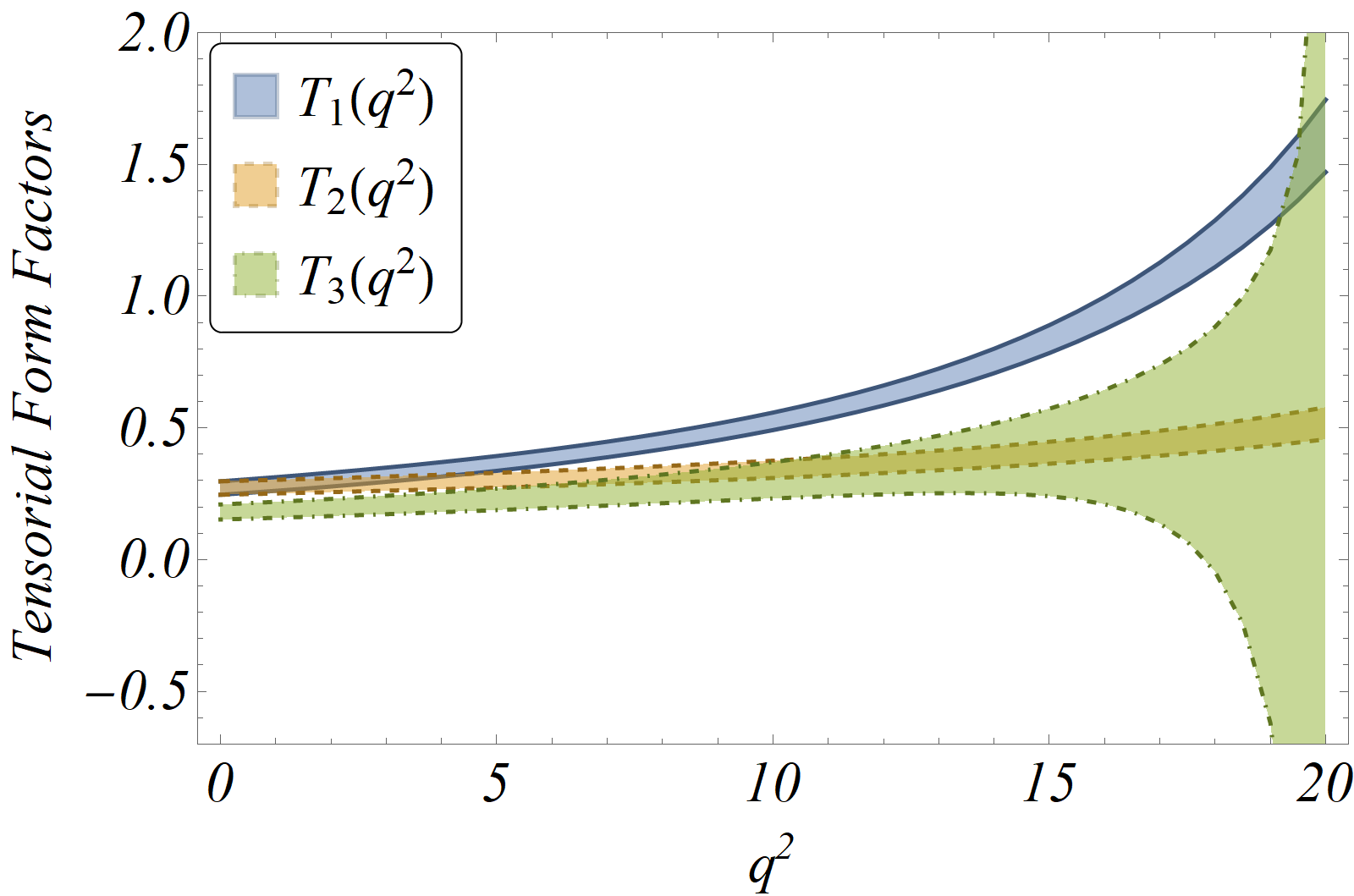}\label{fig:tensorrho}}
	\caption{The comparison of the $q^2$ distributions of the form factors in $B\to \rho \ell\nu$ decays obtained from the fit results in table \ref{tab:fitffrho} for $n=3$ and the one obtained only from the LCSR inputs. The tensor form factors can not be fitted from the experimental inputs that we have used in this analysis.}
	\label{fig:ffactors}
\end{figure}

Using both these fit results, we have estimated the $q^2$ distributions of the differential rates $\Delta \mathcal{B}(B\to \rho\ell\nu)(q^2)$ and compared them with the existing data, which are shown in figure \ref{fig:diffraterho}. We obtain this result using $|V_{ub}| = (3.91 \pm 0.13)\times 10^{-3}$ which is an update on the result reported by~\cite{Biswas:2021qyq} after the inclusion of the LCSR data on $B\to\pi$ form factors from~\cite{Leljak:2021vte}. We find that the result obtained from a fit to all the data points can not accommodate the data points on $ d\mathcal{B}(B^{0(-)} \to\rho^{+(0)} l\nu)/{dq^2}$ from BaBar in the first two bins. Our fit results can accommodate all the data points from Belle without any ambiguity, and are much better when we drop the three data points from Babar. Using the fit results given in table \ref{tab:fitffrho}, we have predicted the $q^2$ distributions of the form factors which are shown in figures\ref{fig:A0}, \ref{fig:A1}, \ref{fig:A2} and \ref{fig:V0} respectively. Although here we have presented our results only for $n=3$, a similar pattern would be observed for $n=2$, albeit with relatively less error. Also, in the same plots, we have compared them with the respective LCSR predictions. As the inputs given in ref.~\cite{Straub:2015ica} is much more precise than the one obtained in ref.~\cite{Gubernari:2018wyi}, our fit results are sensitive to the inputs in \cite{Straub:2015ica}. The LCSR data for the form factors have large errors, and as expected, for high values of $q^2$, the form factors are highly unconstrained. Note that the form factors obtained from a fit to LCSR and the data points on the normalised decay rates are relatively better constrained. In particular, we see that the form factors $A_1(q^2)$ and $A_2(q^2)$ are tightly constrained in all the kinematically allowed $q^2$ regions while for $A_0(q^2)$ and $V(q^2)$ we obtain a precise prediction in the low $q^2$ regions but they are less tightly bound in the high $q^2$ regions. These observations are due to the relative sensitivities of the decay rates to the respective form factors, viz. their susceptibility to $A_1(q^2)$ and $A_2(q^2)$. At the same time, they are relatively less sensitive to $A_0(q^2)$ and $V(q^2)$. For illustration, in the appendix we have presented the predicted values of these form factors at a few different values of $q^2$ in tables \ref{tab:arhffq2} (for all data) and \ref{tab:arhffq2wBabar} (without Babar data), respectively. The $q^2$ distributions of the form factors associated with the tensor current obtained from LCSR are shown in figure \ref{fig:tensorrho}, and the errors are substantial for higher values of $q^2$. We can not constrain these form factors from a fit to the experimental data as of now.

\begin{table}[t]
	\begin{tabular}{|c|c|c|c|c|}
		\hline
		\multirow{3}{*}{\textbf{Observables}} & \multicolumn{4}{c|}{\textbf{SM Predictions}}\\
		\cline{2-5}
		& \multicolumn{2}{c|}{\textbf{Without Babar combined data}} & \multicolumn{2}{c|}{\textbf{With Babar combined data}}  \\
		\cline{2-5}
		&  $\text{n=2}$  &  $\text{n=3}$  &  $\text{n=2}$  &  $\text{n=3}$\\
		\hline
		$\text{R($\rho $)}$  &  $\text{0.5303(57)}$  &  $\text{0.5381(79)}$&  $\text{0.5287(61)}$  &  $\text{0.5352(84)}$\\
		$R_{\tau }^{\tau }\text{($\rho $)}$  &  $\text{0.548(52)}$  &  $\text{0.542(51)}$&  $\text{0.655(57)}$  &  $\text{0.658(58)}$\\
		$R_{\tau }^{\mu }\text{($\rho $)}$  &  $\text{0.00246(23)}$  &  $\text{0.00244(23)}$&  $\text{0.00294(25)}$  &  $\text{0.00295(25)}$\\
		$R_{\tau }^{\tau }\text{($\rho $/$\pi $)}$  &  $\text{1.70(15)}$  &  $\text{1.73(15)}$&  $\text{1.40(10)}$  &  $\text{1.41(10)}$\\
		$R_{\mu }^{\tau }\text{($\rho $/$\pi $)}$  &  $\text{1.163(99)}$  &  $\text{1.182(100)}$&  $\text{0.956(70)}$  &  $\text{0.958(70)}$\\
		$R_{\mu }^{\tau }\text{($\rho $)}$  &  $\text{0.290(27)}$  &  $\text{0.292(27)}$&  $\text{0.346(30)}$  &  $\text{0.352(31)}$\\
		\hline
	\end{tabular}
	\caption{SM predictions for the observables defined in \ref{eq:Rrhoobs}, $R_{\tau }^{\tau }\text{($\rho $/$\pi $)}$ and $R_{\mu }^{\tau }\text{($\rho $/$\pi $)}$ the full allowed kinematic ($0\leq q^2\leq 20.29$) regions corresponding to $n=2$ and $n=3$ parametrizations for the BSZ form factors corresponding to the inclusion and exclusion of the Babar data from~\cite{delAmoSanchez:2010af}.}
	\label{tab:SMpredrho}
\end{table}

Using the results given in table \ref{tab:fitffrho} we have predicted the observables given in eq.~\ref{eq:Rrhoobs} and the following two observables: 
\begin{equation}
R^{\tau}_{\tau}(\rho/\pi) = \frac{\Gamma(B\to\rho\tau\nu)}{\Gamma(B\to\pi\tau\nu)}, \ \ \ R^{\tau}_{\mu}(\rho/\pi) = \frac{\Gamma(B\to\rho\tau\nu)}{\Gamma(B\to\pi\mu\nu)}.
\label{eq:addobsrho}
\end{equation} 
The predictions are provided for both with and without the inclusion of data from the Babar combined analysis. Note that apart from $R(\rho)$, the predictions of all the other observables with and without the data from Babar combined analysis changes, and they are marginally consistent with each other within the given 1-$\sigma$ confidence interval. The observed pattern is due to the reduction of the decay rates $\Gamma(B\to \rho\mu\nu)$ and $\Gamma(B\to \rho\tau\nu)$ in the analysis, which includes Babar data. In $R(\rho)$, the effects cancel in the ratio. One can also understand this from a comparison of the fit results given in table \ref{tab:fitffrho}. The values obtained for $n=3$ should be considered as our final predictions. However, for the purpose of illustration we have presented our results for $n=2$ as well. Note that there is a small shift in the best fit values of the observables depending on whether we truncate the series of eq. \ref{eq:bszexp} at $n=2$ or $n=3$, though they are pretty consistent with each other within the given error bars. The predictions for $R(\rho)$ has an error of $\approx 1\%$ for $n=2$ and that for $n=3$ is roughly $1.5\%$. Therefore, we notice a minimal impact from the coefficients at order $n=3$, which are more uncertain. For the rest of the observables, the predicted errors are relatively large, which are $\approx 10\%$. For these observables, the impacts of the higher-order coefficients are not even noticeable. We need more precise data or inputs from lattice to improve the predictions.  
  
\begin{table}[t]
	\centering
	\begin{tabular}{|c|c|c|c|c|c|c|}
		\hline
		\multirow{4}{*}{\textbf{Observables}} & \multicolumn{6}{c|}{\textbf{Predictions (for $n=3$)}}\\
		\cline{2-7}
		&\multicolumn{2}{c|}{$C_{V_1}$} &\multicolumn{2}{c|}{$C_{V_2}$} &\multicolumn{2}{c|}{$C_P$}  \\
		\cline{2-7}
		& $\text{Sol-1}$  &  $\text{Sol-2}$  &  $\text{Sol-1 }$  &  $\text{Sol-2 }$  &  $\text{Sol-1 }$  &  $\text{Sol-2 }$  \\
		&  -0.20(13) & -1.80(13)  & 0.20(13)   & 1.80(13)   & -0.055(36)  & -0.481(36) \\\hline
		$\text{R($\rho $)}$    &  $\text{0.34(11)}$    &  $\text{0.34(11)}$    &  $\text{0.429(58)}$  &  $\text{1.11(18)}$    &  $\text{0.5303(90)}$    &  $\text{0.4885(77)}$  \\
		$R_{\tau }^{\tau }\text{($\rho $)}$    &  $\text{0.542(51)}$    &  $\text{0.542(51)}$    &  $\text{0.542(51)}$   &  $\text{0.169(33)}$    &  $\text{0.35(12)}$    &  $\text{0.38(13)}$  \\
		$R_{\tau }^{\mu }\text{($\rho $)}$   &  $\text{0.0038(13)}$    &  $\text{0.0038(13)}$    &  $\text{0.00305(49)}$    &  $\text{0.00118(20)}$    &  $\text{0.00247(24)}$    &  $\text{0.00268(25)}$  \\
		$R_{\tau }^{\tau }\text{($\rho $/$\pi $)}$    &  $\text{1.73(15)}$    &  $\text{1.73(15)}$    &  $\text{0.96(34)}$    &  $\text{0.454(45)}$   &  $\text{1.71(15)}$    &  $\text{1.57(14)}$  \\
		$R_{\mu }^{\tau }\text{($\rho $))}$    &  $\text{0.187(63)}$    &  $\text{0.187(63)}$    &  $\text{0.187(63)}$    &  $\text{0.187(63)}$   &  $\text{0.184(65)}$    &  $\text{0.184(65)}$  \\\hline
	\end{tabular}
	\caption{The results of the observables defined in eqns.~\ref{eq:Rrhoobs} and \ref{eq:addobsrho} corresponding to the two NP solutions obtained from fitting to dataset defined in `Fit-3' with NP in $B\to\tau\nu$.}
	\label{tab:NPpredictionsrho}
\end{table}

\begin{figure}[t]
	\centering
	\subfloat[]{\includegraphics[width=0.35\textwidth]{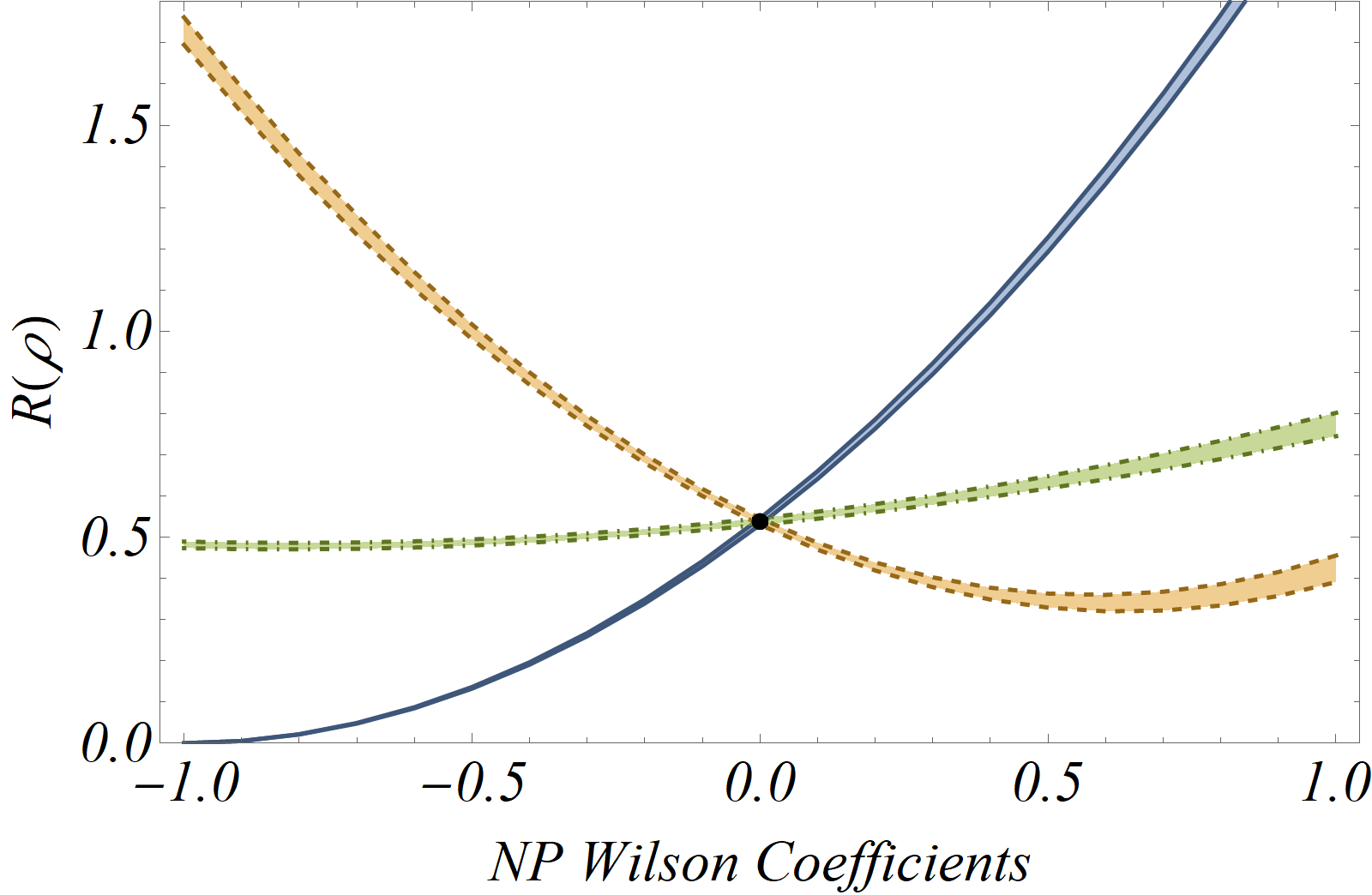}\label{fig:Rrho}}~~~~
	\subfloat[]{\includegraphics[width=0.35\textwidth]{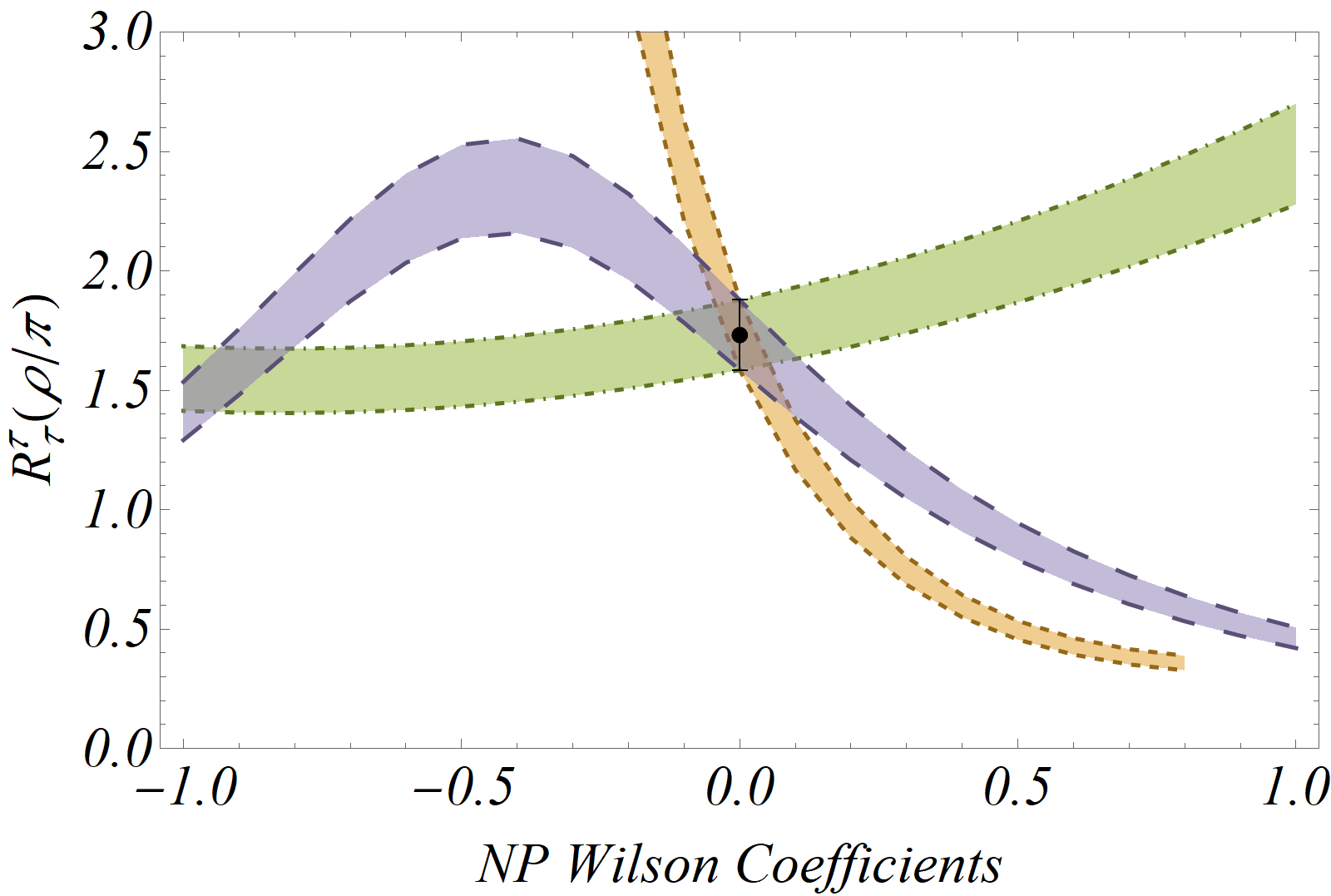}\label{fig:Rrhopitau}}\\
	\subfloat[]{\includegraphics[width=0.35\textwidth]{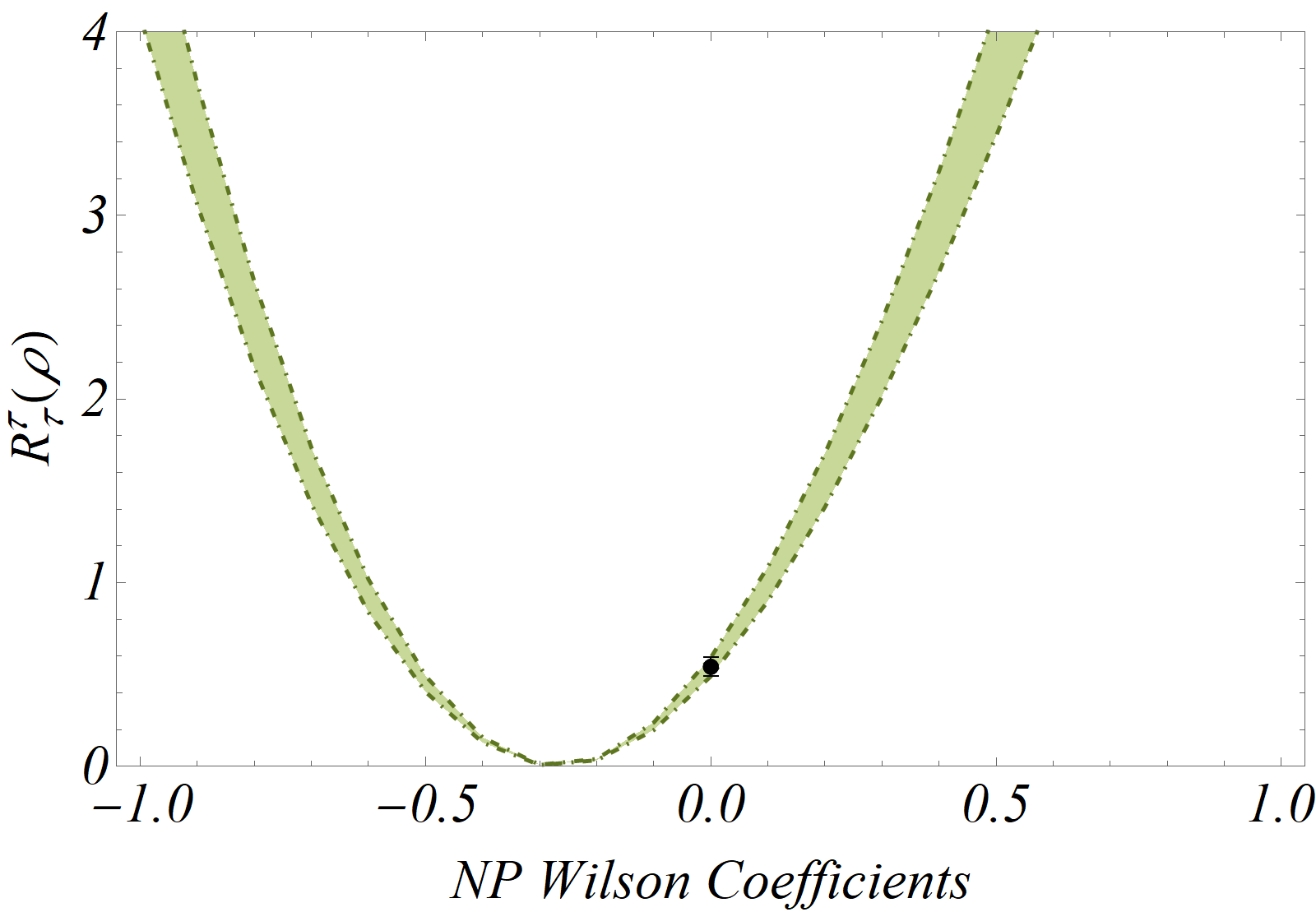}\label{fig:Rtaurho}}~~~~
	\subfloat[]{\includegraphics[width=0.07\textwidth]{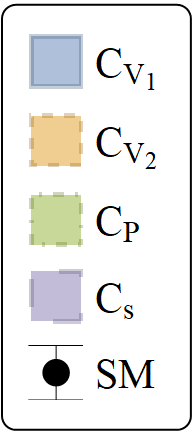}}\\
	\caption{Variations of the observables $R(\rho)$, $R_{\tau }^{\tau }(\rho/\pi)$ and $R_{\tau }^{\mu }(\rho)$ w.r.t. new WC's (real), taken one at a time.}
	\label{fig:obsrhovarnp}
\end{figure}
In table \ref{tab:NPpredictionsrho} we have given the predictions of the observables defined in $B\to \rho$ semileptonic decays in a few NP scenarios as discussed earlier in subsection \ref{subsec:Btopi}. The NP solutions are obtained in the fit scenario `Fit-3', allowing for NP effects in $B\to\tau\nu$ decays. As discussed earlier, sol-2 of $C_{V_2}$ is not allowed by the current data. Depending on the NP scenarios, we note considerable deviations from the respective SM predictions as given in table \ref{tab:SMpredrho} for some observables. In figure \ref{fig:obsrhovarnp}, we have studied the NP sensitivities of the observables $R(\rho)$, $R_{\tau }^{\tau }(\rho )$ and $R_{\tau }^{\tau }(\rho /\pi )$. The NP sensitivities of $R_{\mu }^{\tau }(\rho /\pi )$ and $R(\rho)$ are identical which will be determined by the NP effects in $B\to\rho\tau\nu$, hence we are not discussing them separately. On the other hand, the NP sensitivities of $R_{\tau }^{\mu }(\rho )$ and $R_{\tau }^{\mu }(\pi)$ are similar. Hence we are not repeating the discussion in this subsection. Also, in table \ref{tab:NPpredictionsrho} in the appendix, we have given the predicted values of these observables in different NP scenarios for a few benchmark values of the new WCs.  Certain observations from table \ref{tab:NPpredictionsrho} and figure \ref{fig:obsrhovarnp}:  
\begin{itemize}
	\item For the allowed negative and positive solutions of $C_{V_1}$ and $C_{V_2}$, respectively, we note deviations in $R(\rho)$ from the corresponding SM prediction. In both cases, the predicted value will reduce than the SM. In contrast, the predicted value for $R(\pi)$ will be higher than the respective SM value in the case of $\mathcal{O}_{V_2}$. The observables $R(\pi)$ and $R(\rho)$ will provide complementary informations to distinguish the effects of $\mathcal{O}_{V_1}$ from that of $\mathcal{O}_{V_2}$.  
	
	\item  The obsevable $R_{\tau }^{\tau }(\rho /\pi )$ is insensitive to $C_{V_1}$, while it is sensitive to $C_{V_2}$. In the allowed regions (sol-1) the value of $R_{\tau }^{\tau }(\rho /\pi )$ will reduce than the corresponding SM prediction. 
	
	\item The observables $R(\rho)$ and $R_{\tau }^{\tau }(\rho /\pi )$ are insensitive to the presence of $\mathcal{O}_P$, while $R_{\tau }^{\tau }(\rho )$ is highly sensitive. At present the allowed solutions prefer a value that is lower than the corresponding SM prediction.   
	
	\item The observable $R_{\tau }^{\tau }(\rho /\pi )$ is sensitive to $C_S$ while the other two observables $R(\rho)$ and $R_{\tau }^{\tau }(\rho )$ are not sensitive to $C_S$. For a value of $|C_S| \approx 0.5$, we can observe large deviations in the predicted value w.r.t. the corresponding SM prediction (see table \ref{tab:NPpredictionsrho}).    	
\end{itemize}
Following the above itemised observations we can say that the precise measurements of all these observables will be helpful to distinguish the effects of different new operators.  Note that we have used the fit results without the Babar combined data for this part of the analysis. A similar conclusion holds for the other fit results. 

There will be contributions to these observables from new tensor operators. However, as was shown in figure \ref{fig:tensorrho}, we do not have tight constraints on the respective form factors. The predictions in the full kinematic region will have large errors. Therefore, for this NP scenario, we have presented the predictions in two separate  regions of $q^2$: $0 \le q^2 \le 5$ and $5 \le q^2 \le 10$ (in {\it GeV}$^2$). The respective predictions are given for a few allowed values of $C_T$ which can be seen from table \ref{tab:predrhotensor} in the appendix. All the three observables  are sensitive to $\mathcal{O}_T$ as discussed above.    

\subsection{Angular observables}

\begin{table}[t]
	\begin{tabular}{|c|c|c|c|c|}
		\hline
		\multirow{3}{*}{\textbf{Observables}} & \multicolumn{4}{c|}{\textbf{Values}}\\
		\cline{2-5}
		&\multicolumn{2}{c|}{\textbf{Fit-1}} &\multicolumn{2}{c|}{\textbf{Fit-2}}\\
		\cline{2-5}
		  &  $\text{$\ell=\mu $}$  &  $\text{$\ell=\tau $}$  &  $\text{$\ell=\mu $}$  &  $\text{$\ell=\tau $}$  \\\hline
		$A_{\text{FB}}^l$  &  $\text{0.00484(21)}$  &  $\text{0.2565(31)}$  &  $\text{0.00484(12)}$  &  $\text{0.2564(26)}$  \\
		$P^l$  &  $\text{-0.98679(59)}$  &  $\text{-0.229(21)}$  &  $\text{-0.98683(37)}$  &  $\text{-0.234(19)}$  \\\hline
		\end{tabular}
		\caption{SM values for angular observables related to the semileptonic $B\to\pi$ transitions. The values are provided for both muonic and tauonic final states corresponding to the Fit-1 and Fit-2 scenarios described in the text.}
		\label{tab:angpi}
		\end{table}

\begin{table}
	\begin{tabular}{|c|c|c|c|c|c|c|c|c|}
		\hline
		\multirow{4}{*}{\textbf{Observables}} & \multicolumn{8}{c|}{\textbf{Values}}\\
		\cline{2-9}
		&\multicolumn{4}{c|}{\textbf{Without Babar}} &\multicolumn{4}{c|}{\textbf{With Babar}}\\
		\cline{2-9}
		&\multicolumn{2}{c|}{\textbf{n=2}} &\multicolumn{2}{c|}{\textbf{n=3}}&\multicolumn{2}{c|}{\textbf{n=2}} &\multicolumn{2}{c|}{\textbf{n=3}}\\
		\cline{2-9}
		&  $\text{$\ell=\mu $}$  &  $\text{$\ell=\tau $}$  &  $\text{$\ell=\mu $}$  &  $\text{$\ell=\tau $}$&  $\text{$\ell=\mu $}$  &  $\text{$\ell=\tau $}$  &  $\text{$\ell=\mu $}$  &  $\text{$\ell=\tau $}$  \\\hline
		$A_{\text{FB}}^l$  &  $\text{0.532(60)}$  &  $\text{-0.197(13)}$  &  $\text{0.482(77)}$  &  $\text{-0.215(24)}$&$\text{0.553(64)}$&$\text{-0.202(15)}$&$\text{0.530(80)}$&$\text{-0.212(26)}$  \\
		$P^l$  &  $\text{-0.99088(47)}$  &  $\text{-0.538(13)}$  &  $\text{-0.99120(54)}$  &  $\text{-0.545(20)}$&$\text{-0.99055(49)}$&$\text{-0.524(13)}$&$\text{-0.99070(55)}$&$\text{-0.526(21)}$  \\
		$P^{\rho }$  &  $\text{0.490(16)}$  &  $\text{0.474(14)}$  &  $\text{0.473(29)}$  &  $\text{0.458(28)}$&$\text{0.485(18)}$&$\text{0.471(15)}$&$\text{0.478(30)}$&$\text{0.464(29)}$  \\\hline
	\end{tabular}
	\caption{SM values for angular observables related to the semileptonic $B\to\rho$ transitions for both $n=2$ and $n=3$ BSZ parametrizations of the form factors. The values are provided for both muonic and tauonic final states and corresponding to both the inclusion and exclusion of the Babar data from~\cite{delAmoSanchez:2010af}.}
	\label{tab:angrho}
\end{table}

In the following we define a few angular observables related to decay modes $B \to \pi\ell\nu$ and $B \to \rho\ell\nu$ decays, the detail can be seen from reference \cite{Sakaki:2013bfa}:
\begin{itemize}
\item{ Forward-backward asymmetry is defined as 
	
	\begin{equation}
	{\cal A}_{FB}^{\ell}=\frac{\int_{0}^{1} \frac{d\Gamma}{d\cos\theta} d\cos\theta-\int_{-1}^{0} \frac{d\Gamma}{d\cos\theta}d\cos\theta}
	{\int_{-1}^{1} \frac{d\Gamma}{d\cos\theta} d\cos\theta}, 
	\label{eq10}
	\end{equation}
	where $\theta$ is the angle that $\ell$ ($\mu$ or $\tau$) makes with the $B$ in the rest frame of $\tau{\bar\nu}$. 
}

\item{$\tau$ polarisation asymmetry, 
	\begin{equation}
	P_{\tau} = \frac{\Gamma(\lambda_\tau = 1/2)- \Gamma(\lambda_\tau = -1/2) }{\Gamma(\lambda_\tau = 1/2)+ \Gamma(\lambda_\tau = -1/2) }
	\end{equation}
	where $P_{\tau}$ is the $\tau$-polarisation asymmetries in $B^0\to \pi^+\tau\nu$ and $B^0\to \rho^+\tau\nu$ decays.  }

 \item{$\rho$-longitudinal polarisation ($\rho^+ \to \pi^+\pi^0$),
	
	\begin{equation}
	P_{\rho} = \frac{\Gamma(\lambda_{\rho=0})}{\Gamma(\lambda_{\rho = 0}) 
		+ \Gamma(\lambda_{\rho = 1}) + \Gamma(\lambda_{\rho =-1})},
	\label{eq12}
	\end{equation}}

\item{ The observable $F_{H}^{\ell}$\footnote{The expression for this observable is taken from~\cite{Leljak:2021vte}. The detailed derivation of angular observables in terms of helicity amplitudes for a general $B\to Pl\nu$ decay can be read off from~\cite{Becirevic:2019tpx}. The definition for the observable $F_H$ in terms of the helicity amplitudes is provided in~\cite{Bobeth:2007dw}.}:
	\begin{equation}
	F_H^{\ell} = 1+\frac{2}{3}\frac{1}{\Gamma}\frac{d^2}{d(\cos\theta)^2}\frac{d\Gamma}{d\cos\theta},
	\label{eq13}
	\end{equation}}
\end{itemize}
Here, $\Gamma$ defines the decay rates $\Gamma(B\to M\ell\nu)$. All the above-mentioned observables are potentially sensitive to NP effects, and are expected to be measured with good statistics in the future experimental program like Belle-II. The respective predictions in the SM for $B\to \pi$ and $B\to \rho$ are given in table \ref{tab:angpi} and \ref{tab:angrho}, respectively.

\section{Summary}
We analyse the lattice and the newly available LCSR inputs on the form factors in $\btopilnu$ and $\btorholnu$ ($\ell = \mu$ or $e$) decays along with the experimental data on the decay rates $\Gamma(\btopilnu)$ and $\Gamma(\btorholnu)$ in bins of $q^2$. First, we have extracted the $q^2$ distributions of all the required form factors after fitting the coefficients of the corresponding BSZ. To test the effect of new physics, from the decay rates $\Gamma(B \to M\tau\nu_{\tau})$, $\Gamma(B \to M\ell\nu_{\ell})$, $\Gamma(B \to M\tau\nu_{\tau})$ and $\Gamma(B \to \ell\nu_{\ell})$ (with M = $\pi$ or $\rho$) we have defined some of observables by normalizing one rate by the other. Using the fit results, we have predicted all these observables in the SM. Our best results for a few interesting observables are the following:
\begin{eqnarray}
R(\pi) &=&\frac{\Gamma(B\to\pi\tau\nu)}{\Gamma(B\to\pi\mu\nu)} = 0.673(13),\ \ \ R(\rho) =\frac{\Gamma(B\to\rho\tau\nu)}{\Gamma(B\to\rho\mu\nu)} = 0.493(28) \nn \\
 R_{\mu}^\tau(\pi) &=& \frac{\Gamma(B\to\tau\nu)}{\Gamma(B\to\pi\mu\nu)} = 0.00435(20), \ \ \
 R_{\mu}^\tau(\rho) =\frac{\Gamma(B\to\tau\nu)}{\Gamma(B\to\rho\mu\nu)} = 0.00297(41).
\end{eqnarray}
We have provided the SM predictions of a couple of other observables. These are the most precise predictions available so far. 

 We have studied the NP sensitivities of all the observables we define and have shown how the precise measurements of these observables could be helpful to distinguish types of new interactions beyond the SM. In addition, using the available data on $\mathcal{B}(B\to \tau\nu_{\tau})$ we have constrained some relevant new WCs from a fit to the data on $R_{\mu}^\tau(\pi)$. Though it is not required, sizable new physics contributions are allowed by the data at present. We have also provided predictions of all the observables that we have discussed in different NP scenarios, which can be compared with future measurements.  In addition, we have predicted a few angular observables like forward-backward asymmetries, $\tau$-polarisation asymmetries, $\rho$-longitudinal polarisation, and $F_{H}^{\ell}$ relevant for these decays.       

{\bf Acknowledgments:} This work of S.N. is supported by the Science and Engineering Research Board, Govt. of India, under the grant CRG/2018/001260. We would like to thank Roman Zwicky for useful discussion. Also, AB would like to thank Nicholas James Benoit for pointing out some typos in Table X of our previous version.

\section{Appendix}
We provide some relevant information in the appendix that might help the reader in understanding the results and conclusions stated in this article better. In a nutshell, these are:
\begin{itemize}
	\item Tables~\ref{tab:appffq2} and~\ref{tab:arhffq2} display the values of the $B\to\pi$ and $B\to\rho$ form factors respectively for several $q^2$ (including at $q^2=0$) values within the allowed ranges for the corresponding semileptonic decays. While for $B\to\pi$, these numbers correspond to an $n=3$ BSZ parametrization, for $B\to\rho$ we present the numbers corresponding to both $n=2$ and $n=3$. The kinematic constraints between the different form factors at $q^2=0$ ($f_{+}(0)=f_0(0)$, $T_1(0)=T_2(0)$ etc.) can be verified from these tables. One can also compare these tables to table~\ref{tab:Btopifitresults} (for $B\to\pi$) and table~\ref{tab:fitffrho} for ($B\to\rho$) and verify that the values of the form factors at $q^2=0$ are exactly equal to the value for the first coefficient of the corresponding series, which is a particular feature of the BSZ parametrization for form factors. 
	\item Tables~\ref{tab:nppipred} and~\ref{tab:rhopredictionsNP} exhibit the values of the various observables discussed in the text at a few benchmark values for the NP WC's, corresponding to the $B\to\pi$ and $B\to\rho$ semileptonic transitions respectively. We have provided values for NP WC's within the range $[-0.5,0.5]$, since it is reasonable to expect that the NP, if present will not be large. The corresponding SM values have also been provided. The ``$-$" represent the fact that the contribution from the corresponding WC to the given observable is absent.
	\item The values of the three $B\to\rho$ observables $R(\rho)$, $R_\tau^\tau(\rho/\pi)$ and $R_\tau^\tau(\rho)$ corresponding to the NP WC $C_T$ have been presented in table~\ref{tab:predrhotensor} for two $q^2$ intervals $[0,5]$ and $[0,10]$, along with their respective SM values. As discussed in the text, the tensor form factors that contribute to these observables in the presence of $O_T$ alone cannot be constrained from the binned experimental $B\to\rho l\nu$ data. As such, the only contribution is from LCSR which becomes extremely imprecise at large $q^2$. As such, we demmed it more feasible to present the results corresponding to $C_T$ for the two low $q^2$ bins mentioned above. We have hence refrained from including the contributions due to $C_T$ in figure~\ref{fig:obsrhovarnp} and table~\ref{tab:rhopredictionsNP}.
\end{itemize}
\begin{table}
	\begin{tabular}{|c|c|c|c|c|c|c|c|c|c|}
		\hline
		\multirow{3}{*}{\textbf{$q^2$ Values}} & \multicolumn{9}{c|}{\textbf{Scenarios}}\\
		\cline{2-10}
	&\multicolumn{3}{c|}{\textbf{Fit-1}}	&\multicolumn{3}{c|}{\textbf{Fit-2}} &\multicolumn{3}{c|}{\textbf{Fit-3}}   \\
		\cline{2-10}
	& $f_+(q^2)$ & $f_0(q^2)$& $f_T(q^2)$	& $f_+(q^2)$ &  $f_0(q^2)$ & $f_T(q^2)$&  $f_+(q^2)$ & $f_0(q^2)$& $f_T(q^2)$  \\\hline
		$0$  &  $\text{0.249(14)}$  &  $\text{0.240(14)}$  &  $\text{0.244(15)}$  &  $\text{0.2554(72)}$  &  $\text{0.2457(71)}$  &  $\text{0.250(11)}$  &  $\text{0.260(12)}$  &  $\text{0.250(11)}$  &  $\text{0.253(13)}$  \\
		$5$  &  $\text{0.418(19)}$  &  $\text{0.289(14)}$  &  $\text{0.416(22)}$  &  $\text{0.432(11)}$  &  $\text{0.2953(87)}$  &  $\text{0.426(18)}$  &  $\text{0.434(17)}$  &  $\text{0.299(12)}$  &  $\text{0.428(20)}$  \\
		$9$  &  $\text{0.588(24)}$  &  $\text{0.336(15)}$  &  $\text{0.585(28)}$  &  $\text{0.605(15)}$  &  $\text{0.341(11)}$  &  $\text{0.597(23)}$  &  $\text{0.606(21)}$  &  $\text{0.344(13)}$  &  $\text{0.600(26)}$  \\
		$13$  &  $\text{0.904(31)}$  &  $\text{0.415(15)}$  &  $\text{0.896(36)}$  &  $\text{0.924(21)}$  &  $\text{0.421(13)}$  &  $\text{0.910(32)}$  &  $\text{0.925(28)}$  &  $\text{0.422(14)}$  &  $\text{0.913(34)}$  \\
		$17$  &  $\text{1.630(43)}$  &  $\text{0.570(16)}$  &  $\text{1.603(52)}$  &  $\text{1.644(34)}$  &  $\text{0.573(15)}$  &  $\text{1.611(49)}$  &  $\text{1.649(40)}$  &  $\text{0.574(16)}$  &  $\text{1.618(51)}$  \\
		$21$  &  $\text{4.30(13)}$  &  $\text{0.877(20)}$  &  $\text{4.18(16)}$  &  $\text{4.24(11)}$  &  $\text{0.877(20)}$  &  $\text{4.14(15)}$  &  $\text{4.29(13)}$  &  $\text{0.879(20)}$  &  $\text{4.17(16)}$  \\\hline
	\end{tabular}
	\caption{Values of the different $B\to\pi$ form factors corresponding to a few $q^2$ values within the allowed kinematical region for the $B\to\pi$ semileptonic transitions.}
	\label{tab:appffq2}
\end{table}
\begin{table}[t]
\centering
\begin{tabular}{|c|c|c|c|c|}
\hline
\multirow{2}{*}{\textbf{Scenarios}} & \multirow{2}{*}{\textbf{Values}} & \multicolumn{3}{c|}{\textbf{Observables}}\\
\cline{3-5}
&&$\text{R($\pi $)}$  &  $R_{\tau }^{\tau }\text{($\pi $)}$  &  $R_{\mu }^{\tau }\text{($\pi $)}$  \\\hline
$\text{SM}$  &  $\text{Null}$  &  $\text{0.681(13)}$  &  $\text{0.942(53)}$  &  $\text{0.641(43)}$  \\\hline
$C_{V_1}$  &  $-0.5$  &  $\text{0.1729(39)}$  &  $-$  &  $\text{0.171(14)}$  \\
  &  $-0.05$  &  $\text{0.624(14)}$  &  $-$  &  $\text{0.616(49)}$  \\
  &  $0.05$  &  $\text{0.762(17)}$  &  $-$  &  $\text{0.753(60)}$  \\
  &  $0.5$  &  $\text{1.556(35)}$  &  $-$  &  $\text{1.54(12)}$  \\
\hline
$C_{V_2}$  &  $-0.5$  &  $\text{0.1729(39)}$  &  $\text{8.89(57)}$  &  $\text{1.54(12)}$  \\
  &  $-0.05$  &  $\text{0.624(14)}$  &  $\text{1.207(79)}$  &  $\text{0.753(60)}$  \\
  &  $0.05$  &  $\text{0.762(17)}$  &  $\text{0.809(53)}$  &  $\text{0.616(49)}$  \\
  &  $0.5$  &  $\text{1.556(35)}$  &  $\text{0.1098(72)}$  &  $\text{0.171(14)}$  \\
\hline 
$C_P$  &  $-0.5$  &  $-$  &  $\text{0.741(49)}$  &  $\text{0.512(41)}$  \\
  &  $-0.05$  &  $-$  &  $\text{0.654(43)}$  &  $\text{0.452(36)}$  \\
  &  $0.05$  &  $-$  &  $\text{1.391(91)}$  &  $\text{0.962(77)}$  \\
  &  $0.5$  &  $-$  &  $\text{8.11(53)}$  &  $\text{5.61(45)}$  \\
\hline 
$C_S$  &  $-0.5$  &  $\text{0.5111(98)}$  &  $\text{1.336(91)}$  &  $-$  \\
  &  $-0.05$  &  $\text{0.650(13)}$  &  $\text{1.051(70)}$  &  $-$  \\
  &  $0.05$  &  $\text{0.738(18)}$  &  $\text{0.925(60)}$  &  $-$  \\
  &  $0.5$  &  $\text{1.397(57)}$  &  $\text{0.489(30)}$  &  $-$  \\
\hline 
$C_T$  &  $-0.5$  &  $\text{0.478(21)}$  &  $\text{1.429(87)}$  &  $-$  \\
  &  $-0.05$  &  $\text{0.615(14)}$  &  $\text{1.111(74)}$  &  $-$  \\
  &  $0.05$  &  $\text{0.780(17)}$  &  $\text{0.875(57)}$  &  $-$  \\
  &  $0.5$  &  $\text{2.133(79)}$  &  $\text{0.320(21)}$  &  $-$  \\\hline
\end{tabular}
\caption{NP predictions at differnt values for the corresponding Wilson coefficients (taken one at a time) corresponding to the $B\to\pi$ observables.}
\label{tab:nppipred}
\end{table}

\begin{table}
	\centering
	\begin{tabular}{|c|c|c|c|c|c|c|c|}
		\hline
		\multirow{2}{*}{\textbf{$q^2$ Values}} & \multicolumn{7}{c|}{\textbf{FormFactors}}\\
		\cline{2-8}
		&\textbf{$V(q^2)$}& \textbf{$A_0(q^2)$} &\textbf{$A_1(q^2)$}&\textbf{$A_2(q^2)$}&\textbf{$T_1(q^2)$}&\textbf{$T_2(q^2)$}&\textbf{$T_3(q^2)$}  \\\hline
		$0$  &  $\text{0.310(20)}$  &  $\text{0.309(17)}$  &  $\text{0.247(16)}$  &  $\text{0.225(20)}$  &  $\text{0.278(20)}$  &  $\text{0.278(20)}$  &  $\text{0.202(17)}$  \\
		$4$  &  $\text{0.390(21)}$  &  $\text{0.390(20)}$  &  $\text{0.267(16)}$  &  $\text{0.272(25)}$  &  $\text{0.347(22)}$  &  $\text{0.299(22)}$  &  $\text{0.245(22)}$  \\
		$7$  &  $\text{0.474(22)}$  &  $\text{0.478(24)}$  &  $\text{0.286(17)}$  &  $\text{0.320(32)}$  &  $\text{0.419(24)}$  &  $\text{0.319(24)}$  &  $\text{0.290(27)}$  \\
		$10$  &  $\text{0.595(25)}$  &  $\text{0.602(32)}$  &  $\text{0.308(19)}$  &  $\text{0.384(43)}$  &  $\text{0.519(28)}$  &  $\text{0.344(26)}$  &  $\text{0.352(36)}$  \\
		$13$  &  $\text{0.774(33)}$  &  $\text{0.786(48)}$  &  $\text{0.337(21)}$  &  $\text{0.474(62)}$  &  $\text{0.666(36)}$  &  $\text{0.376(30)}$  &  $\text{0.440(51)}$  \\
		$16$  &  $\text{1.059(55)}$  &  $\text{1.083(81)}$  &  $\text{0.374(25)}$  &  $\text{0.604(95)}$  &  $\text{0.897(53)}$  &  $\text{0.418(36)}$  &  $\text{0.570(76)}$  \\
		$19$  &  $\text{1.56(11)}$  &  $\text{1.61(15)}$  &  $\text{0.425(32)}$  &  $\text{0.81(16)}$  &  $\text{1.298(94)}$  &  $\text{0.479(45)}$  &  $\text{0.78(12)}$\\\hline
	\end{tabular}
	\caption{Values of the different $B\to\rho$ form factors (for n=3) corresponding to a few $q^2$ values within the allowed kinematical region for the $B\to\rho$ semileptonic transitions. The tensorial form factors are constrained by Lattice only and are completely unaltered after the inclusion of the normalized binned experimental data. This result is obtained from a fit that excludes the Babar data.}
	\label{tab:arhffq2}
\end{table}

\begin{table}
	\centering
	\begin{tabular}{|c|c|c|c|c|c|c|c|}
		\hline
		\multirow{2}{*}{\textbf{$q^2$ Values}} & \multicolumn{7}{c|}{\textbf{FormFactors}}\\
		\cline{2-8}
		&\textbf{$V(q^2)$}& \textbf{$A_0(q^2)$} &\textbf{$A_1(q^2)$}&\textbf{$A_2(q^2)$}&\textbf{$T_1(q^2)$}&\textbf{$T_2(q^2)$}&\textbf{$T_3(q^2)$}  \\\hline
		$0$  &  $\text{0.287(19)}$  &  $\text{0.290(16)}$  &  $\text{0.226(15)}$  &  $\text{0.204(19)}$  &  $\text{0.278(20)}$  &  $\text{0.278(20)}$  &  $\text{0.202(17)}$  \\
		$4$  &  $\text{0.362(20)}$  &  $\text{0.367(18)}$  &  $\text{0.243(15)}$  &  $\text{0.244(24)}$  &  $\text{0.347(22)}$  &  $\text{0.299(22)}$  &  $\text{0.245(22)}$  \\
		$7$  &  $\text{0.442(20)}$  &  $\text{0.450(22)}$  &  $\text{0.258(16)}$  &  $\text{0.285(29)}$  &  $\text{0.419(24)}$  &  $\text{0.319(24)}$  &  $\text{0.290(27)}$  \\
		$10$  &  $\text{0.556(22)}$  &  $\text{0.568(30)}$  &  $\text{0.276(17)}$  &  $\text{0.338(40)}$  &  $\text{0.519(28)}$  &  $\text{0.344(26)}$  &  $\text{0.352(36)}$  \\
		$13$  &  $\text{0.726(30)}$  &  $\text{0.743(46)}$  &  $\text{0.299(18)}$  &  $\text{0.412(57)}$  &  $\text{0.666(36)}$  &  $\text{0.376(30)}$  &  $\text{0.440(51)}$  \\
		$16$  &  $\text{0.997(51)}$  &  $\text{1.023(79)}$  &  $\text{0.327(21)}$  &  $\text{0.516(88)}$  &  $\text{0.897(53)}$  &  $\text{0.418(36)}$  &  $\text{0.570(76)}$  \\
		$19$  &  $\text{1.48(10)}$  &  $\text{1.52(15)}$  &  $\text{0.366(26)}$  &  $\text{0.68(14)}$  &  $\text{1.298(94)}$  &  $\text{0.479(45)}$  &  $\text{0.78(12)}$  \\\hline
	\end{tabular}
	\caption{Values of the different $B\to\rho$ form factors (for n=3) corresponding to a few $q^2$ values within the allowed kinematical region for the $B\to\rho$ semileptonic transitions after the inclusion of the Babar data.}
	\label{tab:arhffq2wBabar}
\end{table}

\begin{table}
\centering
\begin{tabular}{|c|c|c|c|c|c|c|c|c|c|}
\hline
\multirow{2}{*}{\textbf{Scenarios}} & \multirow{2}{*}{\textbf{Values}} & \multicolumn{6}{c|}{\textbf{Observables}}\\
\cline{3-8}
&&$\text{R($\rho $)}$  &  $R_{\tau }^{\tau }\text{($\rho $/$\pi $)}$  &  $R_{\mu }^{\tau }\text{($\rho $/$\pi $)}$  &  $R_{\tau }^{\tau }\text{($\rho $)}$  &  $R_{\tau }^{\mu }\text{($\rho $)}$  &  $R_{\mu }^{\tau }\text{($\rho $)}$  \\\hline
$\text{SM}$  &  $\text{Null}$  &  $\text{0.5381(79)}$  &  $\text{1.73(15)}$  &  $\text{1.182(100)}$  &  $\text{0.542(51)}$  &  $\text{0.00244(23)}$  &  $\text{0.292(27)}$  \\\hline
$C_{V_1}$  &  $-0.5$  &  $\text{0.1345(20)}$  &  $-$  &  $\text{0.295(25)}$  &  $-$  &  $\text{0.00974(92)}$  &  $\text{0.0729(68)}$  \\
  &  $-0.05$  &  $\text{0.4856(71)}$  &  $-$  &  $\text{1.067(90)}$  &  $-$  &  $\text{0.00270(25)}$  &  $\text{0.263(25)}$  \\
  &  $0.05$  &  $\text{0.5932(87)}$  &  $-$  &  $\text{1.30(11)}$  &  $-$  &  $\text{0.00221(21)}$  &  $\text{0.322(30)}$  \\
  &  $0.5$  &  $\text{1.211(18)}$  &  $-$  &  $\text{2.66(22)}$  &  $-$  &  $\text{0.00108(10)}$  &  $\text{0.656(61)}$  \\
\hline
$C_{V_2}$  &  $-0.5$  &  $\text{0.999(18)}$  &  $\text{12.9(12)}$  &  $\text{2.19(20)}$  &  $-$  &  $\text{0.00131(12)}$  &  $\text{0.656(61)}$  \\
  &  $-0.05$  &  $\text{0.5720(81)}$  &  $\text{2.04(17)}$  &  $\text{1.26(11)}$  &  $-$  &  $\text{0.00229(22)}$  &  $\text{0.322(30)}$  \\
  &  $0.05$  &  $\text{0.5068(80)}$  &  $\text{1.48(12)}$  &  $\text{1.113(93)}$  &  $-$  &  $\text{0.00259(24)}$  &  $\text{0.263(25)}$  \\
  &  $0.5$  &  $\text{0.346(17)}$  &  $\text{0.495(39)}$  &  $\text{0.761(60)}$  &  $-$  &  $\text{0.00378(31)}$  &  $\text{0.0729(68)}$  \\
\hline 
$C_P$  &  $-0.5$  &  $\text{0.4874(74)}$  &  $\text{1.55(14)}$  &  $\text{1.071(92)}$  &  $\text{0.449(43)}$  &  $\text{0.00269(25)}$  &  $\text{0.219(20)}$  \\
  &  $-0.05$  &  $\text{0.5310(76)}$  &  $\text{1.57(14)}$  &  $\text{1.166(98)}$  &  $\text{0.363(34)}$  &  $\text{0.00247(23)}$  &  $\text{0.193(18)}$  \\
  &  $0.05$  &  $\text{0.5456(82)}$  &  $\text{2.04(17)}$  &  $\text{1.20(10)}$  &  $\text{0.753(71)}$  &  $\text{0.00240(23)}$  &  $\text{0.411(38)}$  \\
  &  $0.5$  &  $\text{0.633(15)}$  &  $\text{2.49(21)}$  &  $\text{1.39(12)}$  &  $\text{3.78(35)}$  &  $\text{0.00207(19)}$  &  $\text{2.40(22)}$  \\
\hline
$C_S$  &  $-0.5$  &  $-$  &  $\text{2.33(20)}$  &  $-$  &  $-$  &  $-$  &  $-$  \\
  &  $-0.05$  &  $-$  &  $\text{1.84(16)}$  &  $-$  &  $-$  &  $-$  &  $-$  \\
  &  $0.05$  &  $-$  &  $\text{1.62(14)}$  &  $-$  &  $-$  &   &   \\
  &  $0.5$  &  $-$  &  $\text{0.867(77)}$  &  $-$  &  $-$  &  $-$  &  $-$  \\\hline
\end{tabular}
\caption{NP predictions at differnt values for the corresponding Wilson coefficients (taken one at a time) corresponding to the $B\to\rho$ observables.}
\label{tab:rhopredictionsNP}
\end{table}
\begin{table}
\begin{tabular}{|c|c|c|c|c|c|c|}
\hline
\multirow{3}{*}{\textbf{$C_T$} Values} & \multicolumn{6}{c|}{\textbf{Observables}}\\
\cline{2-7}
&\multicolumn{2}{c|}{\textbf{$R(\rho)$}} &\multicolumn{2}{c|}{\textbf{$R_\tau^\tau(\rho/\pi)$}} &\multicolumn{2}{c|}{\textbf{$R_\tau^\tau(\rho)$}}  \\
\cline{2-7}
  &  $\text{0$<$}q^2\text{$<$5}$  &  $\text{0$<$}q^2\text{$<$10}$  &  $\text{0$<$}q^2\text{$<$5}$  &  $\text{0$<$}q^2\text{$<$10}$  &  $\text{0$<$}q^2\text{$<$5}$  &  $\text{0$<$}q^2\text{$<$10}$  \\\hline
$0.0$  &  $\text{0.04946(59)}$  &  $\text{0.728(25)}$  &  $\text{0.0399(64)}$  &  $\text{0.588(94)}$  &  $\text{26.4(37)}$  &  $\text{1.80(25)}$ \\\hline
$-0.5$  &  $\text{0.323(26)}$  &  $\text{4.34(35)}$  &  $\text{0.382(57)}$  &  $\text{5.14(79)}$  &  $\text{4.05(54)}$  &  $\text{0.301(41)}$  \\
$-0.05$  &  $\text{0.05568(80)}$  &  $\text{0.823(30)}$  &  $\text{0.0505(81)}$  &  $\text{0.75(12)}$  &  $\text{23.5(32)}$  &  $\text{1.59(22)}$  \\
$0.05$  &  $\text{0.04794(70)}$  &  $\text{0.692(25)}$  &  $\text{0.0343(54)}$  &  $\text{0.496(77)}$  &  $\text{27.3(37)}$  &  $\text{1.89(26)}$  \\
$0.5$  &  $\text{0.246(23)}$  &  $\text{3.04(28)}$  &  $\text{0.0648(98)}$  &  $\text{0.80(12)}$  &  $\text{5.32(69)}$  &  $\text{0.431(52)}$ \\\hline
\end{tabular}
\caption{Values for observables related to the $B\to\rho$ transition for different values of the tensorial NP $C_T$. The tensorial form factors that cannot be constrained from a fit to the $B\to\rho\mu\nu$ binned branching ratios are constrained purely from the LCSR data valid upto $q^2=10$ GeV. The results have hence been presented in two $q^2$ bins [0,5] and [0,10].}
\label{tab:predrhotensor}
\end{table}

\bibliography{ref_AS}

\begin{thebibliography}{29}%
\makeatletter
\providecommand \@ifxundefined [1]{%
 \@ifx{#1\undefined}
}%
\providecommand \@ifnum [1]{%
 \ifnum #1\expandafter \@firstoftwo
 \else \expandafter \@secondoftwo
 \fi
}%
\providecommand \@ifx [1]{%
 \ifx #1\expandafter \@firstoftwo
 \else \expandafter \@secondoftwo
 \fi
}%
\providecommand \natexlab [1]{#1}%
\providecommand \enquote  [1]{``#1''}%
\providecommand \bibnamefont  [1]{#1}%
\providecommand \bibfnamefont [1]{#1}%
\providecommand \citenamefont [1]{#1}%
\providecommand \href@noop [0]{\@secondoftwo}%
\providecommand \href [0]{\begingroup \@sanitize@url \@href}%
\providecommand \@href[1]{\@@startlink{#1}\@@href}%
\providecommand \@@href[1]{\endgroup#1\@@endlink}%
\providecommand \@sanitize@url [0]{\catcode `\\12\catcode `\$12\catcode
  `\&12\catcode `\#12\catcode `\^12\catcode `\_12\catcode `\%12\relax}%
\providecommand \@@startlink[1]{}%
\providecommand \@@endlink[0]{}%
\providecommand \url  [0]{\begingroup\@sanitize@url \@url }%
\providecommand \@url [1]{\endgroup\@href {#1}{\urlprefix }}%
\providecommand \urlprefix  [0]{URL }%
\providecommand \Eprint [0]{\href }%
\providecommand \doibase [0]{https://doi.org/}%
\providecommand \selectlanguage [0]{\@gobble}%
\providecommand \bibinfo  [0]{\@secondoftwo}%
\providecommand \bibfield  [0]{\@secondoftwo}%
\providecommand \translation [1]{[#1]}%
\providecommand \BibitemOpen [0]{}%
\providecommand \bibitemStop [0]{}%
\providecommand \bibitemNoStop [0]{.\EOS\space}%
\providecommand \EOS [0]{\spacefactor3000\relax}%
\providecommand \BibitemShut  [1]{\csname bibitem#1\endcsname}%
\let\auto@bib@innerbib\@empty
\bibitem [{\citenamefont {Amhis}\ \emph {et~al.}(2019)\citenamefont {Amhis}
  \emph {et~al.}}]{Amhis:2019ckw}%
  \BibitemOpen
  \bibfield  {author} {\bibinfo {author} {\bibfnamefont {Y.~S.}\ \bibnamefont
  {Amhis}} \emph {et~al.} (\bibinfo {collaboration} {HFLAV}),\ }\bibfield
  {title} {\bibinfo {title} {{Averages of $b$-hadron, $c$-hadron, and
  $\tau$-lepton properties as of 2018}},\ }\href@noop {} {\  (\bibinfo {year}
  {2019})},\ \Eprint {https://arxiv.org/abs/1909.12524} {arXiv:1909.12524
  [hep-ex]} \BibitemShut {NoStop}%
\bibitem [{\citenamefont {Leljak}\ \emph {et~al.}(2021)\citenamefont {Leljak},
  \citenamefont {Meli\'c},\ and\ \citenamefont {van Dyk}}]{Leljak:2021vte}%
  \BibitemOpen
  \bibfield  {author} {\bibinfo {author} {\bibfnamefont {D.}~\bibnamefont
  {Leljak}}, \bibinfo {author} {\bibfnamefont {B.}~\bibnamefont {Meli\'c}},\
  and\ \bibinfo {author} {\bibfnamefont {D.}~\bibnamefont {van Dyk}},\
  }\bibfield  {title} {\bibinfo {title} {{The $ \overline{B} $
  \textrightarrow{} \ensuremath{\pi} form factors from QCD and their impact on
  |V$_{ub}$|}},\ }\href {https://doi.org/10.1007/JHEP07(2021)036} {\bibfield
  {journal} {\bibinfo  {journal} {JHEP}\ }\textbf {\bibinfo {volume} {07}},\
  \bibinfo {pages} {036}},\ \Eprint {https://arxiv.org/abs/2102.07233}
  {arXiv:2102.07233 [hep-ph]} \BibitemShut {NoStop}%
\bibitem [{\citenamefont {Biswas}\ \emph {et~al.}(2021)\citenamefont {Biswas},
  \citenamefont {Nandi}, \citenamefont {Patra},\ and\ \citenamefont
  {Ray}}]{Biswas:2021qyq}%
  \BibitemOpen
  \bibfield  {author} {\bibinfo {author} {\bibfnamefont {A.}~\bibnamefont
  {Biswas}}, \bibinfo {author} {\bibfnamefont {S.}~\bibnamefont {Nandi}},
  \bibinfo {author} {\bibfnamefont {S.~K.}\ \bibnamefont {Patra}},\ and\
  \bibinfo {author} {\bibfnamefont {I.}~\bibnamefont {Ray}},\ }\bibfield
  {title} {\bibinfo {title} {{A closer look at the extraction of |V$_{ub}$|
  from B \textrightarrow{}
  \ensuremath{\pi}\ensuremath{\ell}\ensuremath{\nu}}},\ }\href
  {https://doi.org/10.1007/JHEP07(2021)082} {\bibfield  {journal} {\bibinfo
  {journal} {JHEP}\ }\textbf {\bibinfo {volume} {07}},\ \bibinfo {pages}
  {082}},\ \Eprint {https://arxiv.org/abs/2103.01809} {arXiv:2103.01809
  [hep-ph]} \BibitemShut {NoStop}%
\bibitem [{\citenamefont {Bernlochner}\ \emph {et~al.}(2021)\citenamefont
  {Bernlochner}, \citenamefont {Prim},\ and\ \citenamefont
  {Robinson}}]{Bernlochner:2021rel}%
  \BibitemOpen
  \bibfield  {author} {\bibinfo {author} {\bibfnamefont {F.~U.}\ \bibnamefont
  {Bernlochner}}, \bibinfo {author} {\bibfnamefont {M.~T.}\ \bibnamefont
  {Prim}},\ and\ \bibinfo {author} {\bibfnamefont {D.~J.}\ \bibnamefont
  {Robinson}},\ }\bibfield  {title} {\bibinfo {title} {{$B \to \rho l \bar \nu$
  and $\omega l \bar \nu$ in and beyond the Standard Model: Improved
  predictions and $|V_{ub}|$}},\ }\href@noop {} {\  (\bibinfo {year} {2021})},\
  \Eprint {https://arxiv.org/abs/2104.05739} {arXiv:2104.05739 [hep-ph]}
  \BibitemShut {NoStop}%
\bibitem [{\citenamefont {Kang}\ \emph {et~al.}(2018)\citenamefont {Kang},
  \citenamefont {Luo}, \citenamefont {Zhang}, \citenamefont {Dai},\ and\
  \citenamefont {Wang}}]{Kang:2018jzg}%
  \BibitemOpen
  \bibfield  {author} {\bibinfo {author} {\bibfnamefont {X.-W.}\ \bibnamefont
  {Kang}}, \bibinfo {author} {\bibfnamefont {T.}~\bibnamefont {Luo}}, \bibinfo
  {author} {\bibfnamefont {Y.}~\bibnamefont {Zhang}}, \bibinfo {author}
  {\bibfnamefont {L.-Y.}\ \bibnamefont {Dai}},\ and\ \bibinfo {author}
  {\bibfnamefont {C.}~\bibnamefont {Wang}},\ }\bibfield  {title} {\bibinfo
  {title} {{Semileptonic $B$ and $B_s$ decays involving scalar and axial-vector
  mesons}},\ }\href {https://doi.org/10.1140/epjc/s10052-018-6385-9} {\bibfield
   {journal} {\bibinfo  {journal} {Eur. Phys. J. C}\ }\textbf {\bibinfo
  {volume} {78}},\ \bibinfo {pages} {909} (\bibinfo {year} {2018})},\ \Eprint
  {https://arxiv.org/abs/1808.02432} {arXiv:1808.02432 [hep-ph]} \BibitemShut
  {NoStop}%
\bibitem [{\citenamefont {Colangelo}\ \emph {et~al.}(2019)\citenamefont
  {Colangelo}, \citenamefont {De~Fazio},\ and\ \citenamefont
  {Loparco}}]{Colangelo:2019axi}%
  \BibitemOpen
  \bibfield  {author} {\bibinfo {author} {\bibfnamefont {P.}~\bibnamefont
  {Colangelo}}, \bibinfo {author} {\bibfnamefont {F.}~\bibnamefont
  {De~Fazio}},\ and\ \bibinfo {author} {\bibfnamefont {F.}~\bibnamefont
  {Loparco}},\ }\bibfield  {title} {\bibinfo {title} {{Probing New Physics with
  $\bar B \to \rho(770) \, \ell^- \bar \nu_\ell$ and $\bar B \to a_1(1260) \,
  \ell^- \bar \nu_\ell$}},\ }\href
  {https://doi.org/10.1103/PhysRevD.100.075037} {\bibfield  {journal} {\bibinfo
   {journal} {Phys. Rev. D}\ }\textbf {\bibinfo {volume} {100}},\ \bibinfo
  {pages} {075037} (\bibinfo {year} {2019})},\ \Eprint
  {https://arxiv.org/abs/1906.07068} {arXiv:1906.07068 [hep-ph]} \BibitemShut
  {NoStop}%
\bibitem [{\citenamefont {Zhang}\ \emph {et~al.}(2021)\citenamefont {Zhang},
  \citenamefont {Kang}, \citenamefont {Guo}, \citenamefont {Dai}, \citenamefont
  {Luo},\ and\ \citenamefont {Wang}}]{Zhang:2020dla}%
  \BibitemOpen
  \bibfield  {author} {\bibinfo {author} {\bibfnamefont {L.}~\bibnamefont
  {Zhang}}, \bibinfo {author} {\bibfnamefont {X.-W.}\ \bibnamefont {Kang}},
  \bibinfo {author} {\bibfnamefont {X.-H.}\ \bibnamefont {Guo}}, \bibinfo
  {author} {\bibfnamefont {L.-Y.}\ \bibnamefont {Dai}}, \bibinfo {author}
  {\bibfnamefont {T.}~\bibnamefont {Luo}},\ and\ \bibinfo {author}
  {\bibfnamefont {C.}~\bibnamefont {Wang}},\ }\bibfield  {title} {\bibinfo
  {title} {{A comprehensive study on the semileptonic decay of heavy flavor
  mesons}},\ }\href {https://doi.org/10.1007/JHEP02(2021)179} {\bibfield
  {journal} {\bibinfo  {journal} {JHEP}\ }\textbf {\bibinfo {volume} {02}},\
  \bibinfo {pages} {179}},\ \Eprint {https://arxiv.org/abs/2012.04417}
  {arXiv:2012.04417 [hep-ph]} \BibitemShut {NoStop}%
\bibitem [{\citenamefont {Be\v{c}irevi\'c}\ \emph {et~al.}(2021)\citenamefont
  {Be\v{c}irevi\'c}, \citenamefont {Jaffredo}, \citenamefont {Pe\~nuelas},\
  and\ \citenamefont {Sumensari}}]{Becirevic:2020rzi}%
  \BibitemOpen
  \bibfield  {author} {\bibinfo {author} {\bibfnamefont {D.}~\bibnamefont
  {Be\v{c}irevi\'c}}, \bibinfo {author} {\bibfnamefont {F.}~\bibnamefont
  {Jaffredo}}, \bibinfo {author} {\bibfnamefont {A.}~\bibnamefont
  {Pe\~nuelas}},\ and\ \bibinfo {author} {\bibfnamefont {O.}~\bibnamefont
  {Sumensari}},\ }\bibfield  {title} {\bibinfo {title} {{New Physics effects in
  leptonic and semileptonic decays}},\ }\href
  {https://doi.org/10.1007/JHEP05(2021)175} {\bibfield  {journal} {\bibinfo
  {journal} {JHEP}\ }\textbf {\bibinfo {volume} {05}},\ \bibinfo {pages}
  {175}},\ \Eprint {https://arxiv.org/abs/2012.09872} {arXiv:2012.09872
  [hep-ph]} \BibitemShut {NoStop}%
\bibitem [{\citenamefont {Fleischer}\ \emph {et~al.}(2021)\citenamefont
  {Fleischer}, \citenamefont {Jaarsma},\ and\ \citenamefont
  {Tetlalmatzi-Xolocotzi}}]{Fleischer:2021yjo}%
  \BibitemOpen
  \bibfield  {author} {\bibinfo {author} {\bibfnamefont {R.}~\bibnamefont
  {Fleischer}}, \bibinfo {author} {\bibfnamefont {R.}~\bibnamefont {Jaarsma}},\
  and\ \bibinfo {author} {\bibfnamefont {G.}~\bibnamefont
  {Tetlalmatzi-Xolocotzi}},\ }\bibfield  {title} {\bibinfo {title} {{Mapping
  out the Space for New Physics with Leptonic and Semileptonic $B_{(c)}$
  Decays}},\ }\href@noop {} {\  (\bibinfo {year} {2021})},\ \Eprint
  {https://arxiv.org/abs/2104.04023} {arXiv:2104.04023 [hep-ph]} \BibitemShut
  {NoStop}%
\bibitem [{hfl()}]{hflavWeb}%
  \BibitemOpen
  \href@noop {} {\bibinfo {title} {$|vub|$ from exclusive seimileptonic b
  decays}},\ \bibinfo {howpublished}
  {\url{https://hflav-eos.web.cern.ch/hflav-eos/semi/summer16/html/ExclusiveVub/exclPilnu.html}}\BibitemShut
  {NoStop}%
\bibitem [{\citenamefont {Jaiswal}\ \emph {et~al.}(2017)\citenamefont
  {Jaiswal}, \citenamefont {Nandi},\ and\ \citenamefont
  {Patra}}]{Jaiswal:2017rve}%
  \BibitemOpen
  \bibfield  {author} {\bibinfo {author} {\bibfnamefont {S.}~\bibnamefont
  {Jaiswal}}, \bibinfo {author} {\bibfnamefont {S.}~\bibnamefont {Nandi}},\
  and\ \bibinfo {author} {\bibfnamefont {S.~K.}\ \bibnamefont {Patra}},\
  }\bibfield  {title} {\bibinfo {title} {{Extraction of $|V_{cb}|$ from $B\to
  D^{(*)}\ell\nu_\ell$ and the Standard Model predictions of $R(D^{(*)})$}},\
  }\href {https://doi.org/10.1007/JHEP12(2017)060} {\bibfield  {journal}
  {\bibinfo  {journal} {JHEP}\ }\textbf {\bibinfo {volume} {12}},\ \bibinfo
  {pages} {060}},\ \Eprint {https://arxiv.org/abs/1707.09977} {arXiv:1707.09977
  [hep-ph]} \BibitemShut {NoStop}%
\bibitem [{\citenamefont {Jaiswal}\ \emph {et~al.}(2020)\citenamefont
  {Jaiswal}, \citenamefont {Nandi},\ and\ \citenamefont
  {Patra}}]{Jaiswal:2020wer}%
  \BibitemOpen
  \bibfield  {author} {\bibinfo {author} {\bibfnamefont {S.}~\bibnamefont
  {Jaiswal}}, \bibinfo {author} {\bibfnamefont {S.}~\bibnamefont {Nandi}},\
  and\ \bibinfo {author} {\bibfnamefont {S.~K.}\ \bibnamefont {Patra}},\
  }\bibfield  {title} {\bibinfo {title} {{Updates on extraction of |V$_{cb}$|
  and SM prediction of R(D*) in $B\to D^{*}\ell\nu_\ell$ decays}},\ }\href
  {https://doi.org/10.1007/JHEP06(2020)165} {\bibfield  {journal} {\bibinfo
  {journal} {JHEP}\ }\textbf {\bibinfo {volume} {06}},\ \bibinfo {pages}
  {165}},\ \Eprint {https://arxiv.org/abs/2002.05726} {arXiv:2002.05726
  [hep-ph]} \BibitemShut {NoStop}%
\bibitem [{\citenamefont {Banelli}\ \emph {et~al.}(2018)\citenamefont
  {Banelli}, \citenamefont {Fleischer}, \citenamefont {Jaarsma},\ and\
  \citenamefont {Tetlalmatzi-Xolocotzi}}]{Banelli:2018fnx}%
  \BibitemOpen
  \bibfield  {author} {\bibinfo {author} {\bibfnamefont {G.}~\bibnamefont
  {Banelli}}, \bibinfo {author} {\bibfnamefont {R.}~\bibnamefont {Fleischer}},
  \bibinfo {author} {\bibfnamefont {R.}~\bibnamefont {Jaarsma}},\ and\ \bibinfo
  {author} {\bibfnamefont {G.}~\bibnamefont {Tetlalmatzi-Xolocotzi}},\
  }\bibfield  {title} {\bibinfo {title} {{Decoding (Pseudo)-Scalar Operators in
  Leptonic and Semileptonic $B$ Decays}},\ }\href
  {https://doi.org/10.1140/epjc/s10052-018-6393-9} {\bibfield  {journal}
  {\bibinfo  {journal} {Eur. Phys. J. C}\ }\textbf {\bibinfo {volume} {78}},\
  \bibinfo {pages} {911} (\bibinfo {year} {2018})},\ \Eprint
  {https://arxiv.org/abs/1809.09051} {arXiv:1809.09051 [hep-ph]} \BibitemShut
  {NoStop}%
\bibitem [{\citenamefont {Flynn}\ \emph {et~al.}(2015)\citenamefont {Flynn},
  \citenamefont {Izubuchi}, \citenamefont {Kawanai}, \citenamefont {Lehner},
  \citenamefont {Soni}, \citenamefont {Van~de Water},\ and\ \citenamefont
  {Witzel}}]{Flynn:2015mha}%
  \BibitemOpen
  \bibfield  {author} {\bibinfo {author} {\bibfnamefont {J.~M.}\ \bibnamefont
  {Flynn}}, \bibinfo {author} {\bibfnamefont {T.}~\bibnamefont {Izubuchi}},
  \bibinfo {author} {\bibfnamefont {T.}~\bibnamefont {Kawanai}}, \bibinfo
  {author} {\bibfnamefont {C.}~\bibnamefont {Lehner}}, \bibinfo {author}
  {\bibfnamefont {A.}~\bibnamefont {Soni}}, \bibinfo {author} {\bibfnamefont
  {R.~S.}\ \bibnamefont {Van~de Water}},\ and\ \bibinfo {author} {\bibfnamefont
  {O.}~\bibnamefont {Witzel}},\ }\bibfield  {title} {\bibinfo {title} {{$B \to
  \pi \ell \nu$ and $B_s \to K \ell \nu$ form factors and $|V_{ub}|$ from
  2+1-flavor lattice QCD with domain-wall light quarks and relativistic heavy
  quarks}},\ }\href {https://doi.org/10.1103/PhysRevD.91.074510} {\bibfield
  {journal} {\bibinfo  {journal} {Phys. Rev. D}\ }\textbf {\bibinfo {volume}
  {91}},\ \bibinfo {pages} {074510} (\bibinfo {year} {2015})},\ \Eprint
  {https://arxiv.org/abs/1501.05373} {arXiv:1501.05373 [hep-lat]} \BibitemShut
  {NoStop}%
\bibitem [{\citenamefont {Bailey}\ \emph
  {et~al.}(2015{\natexlab{a}})\citenamefont {Bailey} \emph
  {et~al.}}]{Lattice:2015tia}%
  \BibitemOpen
  \bibfield  {author} {\bibinfo {author} {\bibfnamefont {J.~A.}\ \bibnamefont
  {Bailey}} \emph {et~al.} (\bibinfo {collaboration} {Fermilab Lattice,
  MILC}),\ }\bibfield  {title} {\bibinfo {title} {{$|V_{ub}|$ from
  $B\to\pi\ell\nu$ decays and (2+1)-flavor lattice QCD}},\ }\href
  {https://doi.org/10.1103/PhysRevD.92.014024} {\bibfield  {journal} {\bibinfo
  {journal} {Phys. Rev. D}\ }\textbf {\bibinfo {volume} {92}},\ \bibinfo
  {pages} {014024} (\bibinfo {year} {2015}{\natexlab{a}})},\ \Eprint
  {https://arxiv.org/abs/1503.07839} {arXiv:1503.07839 [hep-lat]} \BibitemShut
  {NoStop}%
\bibitem [{\citenamefont {Gubernari}\ \emph {et~al.}(2019)\citenamefont
  {Gubernari}, \citenamefont {Kokulu},\ and\ \citenamefont {van
  Dyk}}]{Gubernari:2018wyi}%
  \BibitemOpen
  \bibfield  {author} {\bibinfo {author} {\bibfnamefont {N.}~\bibnamefont
  {Gubernari}}, \bibinfo {author} {\bibfnamefont {A.}~\bibnamefont {Kokulu}},\
  and\ \bibinfo {author} {\bibfnamefont {D.}~\bibnamefont {van Dyk}},\
  }\bibfield  {title} {\bibinfo {title} {{$B\to P$ and $B\to V$ Form Factors
  from $B$-Meson Light-Cone Sum Rules beyond Leading Twist}},\ }\href
  {https://doi.org/10.1007/JHEP01(2019)150} {\bibfield  {journal} {\bibinfo
  {journal} {JHEP}\ }\textbf {\bibinfo {volume} {01}},\ \bibinfo {pages}
  {150}},\ \Eprint {https://arxiv.org/abs/1811.00983} {arXiv:1811.00983
  [hep-ph]} \BibitemShut {NoStop}%
\bibitem [{\citenamefont {Bailey}\ \emph
  {et~al.}(2015{\natexlab{b}})\citenamefont {Bailey} \emph
  {et~al.}}]{Bailey:2015nbd}%
  \BibitemOpen
  \bibfield  {author} {\bibinfo {author} {\bibfnamefont {J.~A.}\ \bibnamefont
  {Bailey}} \emph {et~al.} (\bibinfo {collaboration} {Fermilab Lattice,
  MILC}),\ }\bibfield  {title} {\bibinfo {title} {{$B\to\pi\ell\ell$ form
  factors for new-physics searches from lattice QCD}},\ }\href
  {https://doi.org/10.1103/PhysRevLett.115.152002} {\bibfield  {journal}
  {\bibinfo  {journal} {Phys. Rev. Lett.}\ }\textbf {\bibinfo {volume} {115}},\
  \bibinfo {pages} {152002} (\bibinfo {year} {2015}{\natexlab{b}})},\ \Eprint
  {https://arxiv.org/abs/1507.01618} {arXiv:1507.01618 [hep-ph]} \BibitemShut
  {NoStop}%
\bibitem [{\citenamefont {Ha}\ \emph {et~al.}(2011)\citenamefont {Ha} \emph
  {et~al.}}]{Ha:2010rf}%
  \BibitemOpen
  \bibfield  {author} {\bibinfo {author} {\bibfnamefont {H.}~\bibnamefont {Ha}}
  \emph {et~al.} (\bibinfo {collaboration} {Belle}),\ }\bibfield  {title}
  {\bibinfo {title} {{Measurement of the decay $B^0\to\pi^-\ell^+\nu$ and
  determination of $|V_{ub}|$}},\ }\href
  {https://doi.org/10.1103/PhysRevD.83.071101} {\bibfield  {journal} {\bibinfo
  {journal} {Phys. Rev. D}\ }\textbf {\bibinfo {volume} {83}},\ \bibinfo
  {pages} {071101} (\bibinfo {year} {2011})},\ \Eprint
  {https://arxiv.org/abs/1012.0090} {arXiv:1012.0090 [hep-ex]} \BibitemShut
  {NoStop}%
\bibitem [{\citenamefont {Lees}\ \emph {et~al.}(2012)\citenamefont {Lees} \emph
  {et~al.}}]{Lees:2012vv}%
  \BibitemOpen
  \bibfield  {author} {\bibinfo {author} {\bibfnamefont {J.}~\bibnamefont
  {Lees}} \emph {et~al.} (\bibinfo {collaboration} {BaBar}),\ }\bibfield
  {title} {\bibinfo {title} {{Branching fraction and form-factor shape
  measurements of exclusive charmless semileptonic B decays, and determination
  of $|V_{ub}|$}},\ }\href {https://doi.org/10.1103/PhysRevD.86.092004}
  {\bibfield  {journal} {\bibinfo  {journal} {Phys. Rev. D}\ }\textbf {\bibinfo
  {volume} {86}},\ \bibinfo {pages} {092004} (\bibinfo {year} {2012})},\
  \Eprint {https://arxiv.org/abs/1208.1253} {arXiv:1208.1253 [hep-ex]}
  \BibitemShut {NoStop}%
\bibitem [{\citenamefont {Sibidanov}\ \emph {et~al.}(2013)\citenamefont
  {Sibidanov} \emph {et~al.}}]{Sibidanov:2013rkk}%
  \BibitemOpen
  \bibfield  {author} {\bibinfo {author} {\bibfnamefont {A.}~\bibnamefont
  {Sibidanov}} \emph {et~al.} (\bibinfo {collaboration} {Belle}),\ }\bibfield
  {title} {\bibinfo {title} {{Study of Exclusive $B \to X_u \ell \nu$ Decays
  and Extraction of $\|V_{ub}\|$ using Full Reconstruction Tagging at the Belle
  Experiment}},\ }\href {https://doi.org/10.1103/PhysRevD.88.032005} {\bibfield
   {journal} {\bibinfo  {journal} {Phys. Rev. D}\ }\textbf {\bibinfo {volume}
  {88}},\ \bibinfo {pages} {032005} (\bibinfo {year} {2013})},\ \Eprint
  {https://arxiv.org/abs/1306.2781} {arXiv:1306.2781 [hep-ex]} \BibitemShut
  {NoStop}%
\bibitem [{\citenamefont {Bharucha}\ \emph {et~al.}(2016)\citenamefont
  {Bharucha}, \citenamefont {Straub},\ and\ \citenamefont
  {Zwicky}}]{Straub:2015ica}%
  \BibitemOpen
  \bibfield  {author} {\bibinfo {author} {\bibfnamefont {A.}~\bibnamefont
  {Bharucha}}, \bibinfo {author} {\bibfnamefont {D.~M.}\ \bibnamefont
  {Straub}},\ and\ \bibinfo {author} {\bibfnamefont {R.}~\bibnamefont
  {Zwicky}},\ }\bibfield  {title} {\bibinfo {title} {{$B\to V\ell^+\ell^-$ in
  the Standard Model from light-cone sum rules}},\ }\href
  {https://doi.org/10.1007/JHEP08(2016)098} {\bibfield  {journal} {\bibinfo
  {journal} {JHEP}\ }\textbf {\bibinfo {volume} {08}},\ \bibinfo {pages}
  {098}},\ \Eprint {https://arxiv.org/abs/1503.05534} {arXiv:1503.05534
  [hep-ph]} \BibitemShut {NoStop}%
\bibitem [{pdg({\natexlab{a}})}]{pdgrev}%
  \BibitemOpen
  \href@noop {} {\bibinfo {title} {Ckmfitter global fit results as of summer
  19}},\ \bibinfo {howpublished}
  {\url{https://ckmfitter.in2p3.fr/www/results/plots_summer19/num/ckmEval_results_summer19.html}}
  ({\natexlab{a}})\BibitemShut {NoStop}%
\bibitem [{\citenamefont {Aoki}\ \emph {et~al.}(2020)\citenamefont {Aoki} \emph
  {et~al.}}]{FlavourLatticeAveragingGroup:2019iem}%
  \BibitemOpen
  \bibfield  {author} {\bibinfo {author} {\bibfnamefont {S.}~\bibnamefont
  {Aoki}} \emph {et~al.} (\bibinfo {collaboration} {Flavour Lattice Averaging
  Group}),\ }\bibfield  {title} {\bibinfo {title} {{FLAG Review 2019: Flavour
  Lattice Averaging Group (FLAG)}},\ }\href
  {https://doi.org/10.1140/epjc/s10052-019-7354-7} {\bibfield  {journal}
  {\bibinfo  {journal} {Eur. Phys. J. C}\ }\textbf {\bibinfo {volume} {80}},\
  \bibinfo {pages} {113} (\bibinfo {year} {2020})},\ \Eprint
  {https://arxiv.org/abs/1902.08191} {arXiv:1902.08191 [hep-lat]} \BibitemShut
  {NoStop}%
\bibitem [{pdg({\natexlab{b}})}]{pdg}%
  \BibitemOpen
  \href@noop {} {\bibinfo {title} {Semileptonic b-hadron decays, determination
  of vcb, vub}},\ \bibinfo {howpublished}
  {\url{https://pdg.lbl.gov/2020/reviews/rpp2020-rev-vcb-vub.pdf}}
  ({\natexlab{b}})\BibitemShut {NoStop}%
\bibitem [{\citenamefont {Amhis}\ \emph {et~al.}(2021)\citenamefont {Amhis}
  \emph {et~al.}}]{HFLAV:2019otj}%
  \BibitemOpen
  \bibfield  {author} {\bibinfo {author} {\bibfnamefont {Y.~S.}\ \bibnamefont
  {Amhis}} \emph {et~al.} (\bibinfo {collaboration} {HFLAV}),\ }\bibfield
  {title} {\bibinfo {title} {{Averages of b-hadron, c-hadron, and $\tau
  $-lepton properties as of 2018}},\ }\href
  {https://doi.org/10.1140/epjc/s10052-020-8156-7} {\bibfield  {journal}
  {\bibinfo  {journal} {Eur. Phys. J. C}\ }\textbf {\bibinfo {volume} {81}},\
  \bibinfo {pages} {226} (\bibinfo {year} {2021})},\ \Eprint
  {https://arxiv.org/abs/1909.12524} {arXiv:1909.12524 [hep-ex]} \BibitemShut
  {NoStop}%
\bibitem [{\citenamefont {del Amo~Sanchez}\ \emph {et~al.}(2011)\citenamefont
  {del Amo~Sanchez} \emph {et~al.}}]{delAmoSanchez:2010af}%
  \BibitemOpen
  \bibfield  {author} {\bibinfo {author} {\bibfnamefont {P.}~\bibnamefont {del
  Amo~Sanchez}} \emph {et~al.} (\bibinfo {collaboration} {BaBar}),\ }\bibfield
  {title} {\bibinfo {title} {{Study of $B \to \pi \ell \nu$ and $B \to \rho
  \ell \nu$ Decays and Determination of $|V_{ub}|$}},\ }\href
  {https://doi.org/10.1103/PhysRevD.83.032007} {\bibfield  {journal} {\bibinfo
  {journal} {Phys. Rev. D}\ }\textbf {\bibinfo {volume} {83}},\ \bibinfo
  {pages} {032007} (\bibinfo {year} {2011})},\ \Eprint
  {https://arxiv.org/abs/1005.3288} {arXiv:1005.3288 [hep-ex]} \BibitemShut
  {NoStop}%
\bibitem [{\citenamefont {Sakaki}\ \emph {et~al.}(2013)\citenamefont {Sakaki},
  \citenamefont {Tanaka}, \citenamefont {Tayduganov},\ and\ \citenamefont
  {Watanabe}}]{Sakaki:2013bfa}%
  \BibitemOpen
  \bibfield  {author} {\bibinfo {author} {\bibfnamefont {Y.}~\bibnamefont
  {Sakaki}}, \bibinfo {author} {\bibfnamefont {M.}~\bibnamefont {Tanaka}},
  \bibinfo {author} {\bibfnamefont {A.}~\bibnamefont {Tayduganov}},\ and\
  \bibinfo {author} {\bibfnamefont {R.}~\bibnamefont {Watanabe}},\ }\bibfield
  {title} {\bibinfo {title} {{Testing leptoquark models in $\bar B \to D^{(*)}
  \tau \bar\nu$}},\ }\href {https://doi.org/10.1103/PhysRevD.88.094012}
  {\bibfield  {journal} {\bibinfo  {journal} {Phys. Rev. D}\ }\textbf {\bibinfo
  {volume} {88}},\ \bibinfo {pages} {094012} (\bibinfo {year} {2013})},\
  \Eprint {https://arxiv.org/abs/1309.0301} {arXiv:1309.0301 [hep-ph]}
  \BibitemShut {NoStop}%
\bibitem [{\citenamefont {Be\v{c}irevi\'c}\ \emph {et~al.}(2019)\citenamefont
  {Be\v{c}irevi\'c}, \citenamefont {Fedele}, \citenamefont
  {Ni\v{s}and\v{z}i\'c},\ and\ \citenamefont {Tayduganov}}]{Becirevic:2019tpx}%
  \BibitemOpen
  \bibfield  {author} {\bibinfo {author} {\bibfnamefont {D.}~\bibnamefont
  {Be\v{c}irevi\'c}}, \bibinfo {author} {\bibfnamefont {M.}~\bibnamefont
  {Fedele}}, \bibinfo {author} {\bibfnamefont {I.}~\bibnamefont
  {Ni\v{s}and\v{z}i\'c}},\ and\ \bibinfo {author} {\bibfnamefont
  {A.}~\bibnamefont {Tayduganov}},\ }\bibfield  {title} {\bibinfo {title}
  {{Lepton Flavor Universality tests through angular observables of
  $\overline{B}\to D^{(\ast)}\ell\overline{\nu}$ decay modes}},\ }\href@noop {}
  {\  (\bibinfo {year} {2019})},\ \Eprint {https://arxiv.org/abs/1907.02257}
  {arXiv:1907.02257 [hep-ph]} \BibitemShut {NoStop}%
\bibitem [{\citenamefont {Bobeth}\ \emph {et~al.}(2007)\citenamefont {Bobeth},
  \citenamefont {Hiller},\ and\ \citenamefont {Piranishvili}}]{Bobeth:2007dw}%
  \BibitemOpen
  \bibfield  {author} {\bibinfo {author} {\bibfnamefont {C.}~\bibnamefont
  {Bobeth}}, \bibinfo {author} {\bibfnamefont {G.}~\bibnamefont {Hiller}},\
  and\ \bibinfo {author} {\bibfnamefont {G.}~\bibnamefont {Piranishvili}},\
  }\bibfield  {title} {\bibinfo {title} {{Angular distributions of $\bar{B} \to
  \bar{K} \ell^+\ell^-$ decays}},\ }\href
  {https://doi.org/10.1088/1126-6708/2007/12/040} {\bibfield  {journal}
  {\bibinfo  {journal} {JHEP}\ }\textbf {\bibinfo {volume} {12}},\ \bibinfo
  {pages} {040}},\ \Eprint {https://arxiv.org/abs/0709.4174} {arXiv:0709.4174
  [hep-ph]} \BibitemShut {NoStop}%
\end{thebibliography}%

\end{document}